\documentclass[11pt]{article}
\usepackage{mathtools}
\usepackage{amsmath}
\usepackage{amsfonts}
\usepackage{dsfont}
\usepackage{amssymb}
\usepackage{graphicx}
\usepackage{chicago}
\usepackage[all]{xy}
\usepackage{xcolor}
\usepackage{multirow}
\usepackage{hyperref}
\usepackage[nolists]{endfloat}

\usepackage{cleveref}
\usepackage{tikz}
\DeclareRobustCommand\fulllwd  {\tikz[baseline=-0.6ex]\draw[line width=0.5mm] (0,0)--(0.5,0);}
\DeclareRobustCommand\full  {\tikz[baseline=-0.6ex]\draw[thick] (0,0)--(0.5,0);}

\DeclareRobustCommand\dashed{\tikz[baseline=-0.6ex]\draw[thick,dashed] (0,0)--(0.54,0);}

\DeclareMathAlphabet{\mathcalligra}{T1}{calligra}{m}{k}
\DeclareMathAlphabet{\mathpzc}{OT1}{pzc}{m}{it}

\newtheorem{Remark}{Remark}

\begin{document}

\begin{titlepage}

\begin{center}
\bf\LARGE 

Pareto models for risk management
\par\end{center}

\bigskip

\begin{center}
\Large by 
\par\end{center}

\renewcommand*{\thefootnote}{\fnsymbol{footnote}}
\begin{center}
\Large
\bigskip
\textbf{Arthur Charpentier}\\[1ex]
\large
Universit\' e du Qu\'ebec \`a Montr\'eal (UQAM)\\
201, avenue du Pr\'esident-Kennedy, \\
Montr\'eal (Qu\'ebec), Canada H2X 3Y7
\\ arthur.charpentier{@}uqam.ca

\bigskip
{ and}

\bigskip

\Large
\textbf{Emmanuel Flachaire}\\[1ex]
\large
Aix-Marseille Universit\'e  \\
AMSE, CNRS and EHESS, \\ 
   5 bd Maurice Bourdet, \\ 13001, 
   Marseille, France, \\ emmanuel.flachaire{@}univ-amu.fr

\par\end{center}

\setcounter{footnote}{0}

\vfill
\begin{center}
\large 
December 2019
\par\end{center}

\vfill

\end{titlepage}

\begin{abstract}
The Pareto model is very popular in risk management, since simple analytical formulas can be derived for financial downside risk measures (Value-at-Risk, Expected Shortfall) or reinsurance premiums and related quantities (Large Claim Index, Return Period). Nevertheless, in practice, distributions are (strictly) Pareto only in the tails, above (possible very) large threshold. Therefore, it could be interesting to take into account second order behavior to provide a better fit. In this article, we present how to go from a strict Pareto model to Pareto-type distributions. We discuss inference, and derive formulas for various measures and indices, and finally provide applications on insurance losses and financial risks.




\vfill
\end{abstract}


\section{Introduction}

The Gaussian distribution is one of the most popular distribution used initially to model observation errors. It can be used to model individual measurements that are (somehow) centered, such as the height, or the weight of individuals. But not all data exhibit such a property, where the distribution has a peak around a typical value (the median or the mean), such as the distribution of the population among cities, or the distribution of wealth among individuals. Those distribution are known to be right-skewed, where most observations are bulked with small values, a small proportion can reach much higher values than the median (or the mean), leading to long tail on the right of the distribution. For instance, \citeN{Pareto} introduced the following model for the distribution of income: let $p(x)dx$ denote the proportion of individuals with income between $x$ and $x+dx$; if the histogram is a straight line on the log-log scales, then $\log p(x)=a-b\log(x)$ (since the slope is clearly always negative), or equivalently, $p(x)=c x^{-b}$. Such distributions are said to have a {\em power law}, or a Pareto distribution since it was introduced by the economist Vilfredo Pareto (see \citeN{CharpentierFlachaire} for applications to model income distributions). The constant $b$ might be called the exponent of the power law (especially in the context of networks or in physical applications), but in economic application, the exponent will be the one of the survival distribution ${S(x)=\int_x^{\infty} p(y)dy}$, which is also a power function, with index $1+b$.

The Gaussian distribution is a natural candidate when focusing on the center of the distribution, on average values, because of the central limit theorem: the Gaussian distribution is stable by summing, and averaging. If $\{X_1,\cdots,X_n\}$ are independent Gaussian variables, so is $X_1+\cdots+X_n$ (and any linear transformation).
As mentioned in \citeN{Jenssen:06} and \citeN{Gabaix}, the power distribution satisfies similar interesting properties: if $X_1$ is a power distribution with exponent $b_1$, independent of $X_2$, another power distribution with exponent $b_2$, with $b_2\geq b_1$, then $X_1+X_2$, $X_1\cdot X_2$, or $\max\{X_1,X_2\}$ are also power distributed, with exponent $b_1$. And those properties of invariance and stability are essential in stochastic modeling. As mentioned in \citeN{Schumpeter}, in an obituary article about Vilfredo Pareto, ``{\em few if any economists seem to have realized the possibilities that such invariants hold for the future of our science. In particular, nobody seems to have realized that the hunt for, and the interpretation of, invariants of this type might lay the foundations for an entirely novel type of theory}''. If several applications can be found on proportional (random) growth theory (as discussed in \shortciteN{Gabaix}, with connections to recursive equations), or on matching and networks, an important application of this Pareto model is risk management.

Insurers and actuaries have used Pareto distribution to model large losses since \citeANP{Hagstroem:25} (\citeyearNP{Hagstroem:25}, \citeyearNP{Hagstroem:60}). \citeN{BaHa:74} suggested to use such a distribution in a life-insurance context (to model the remaining lifetimes for very old people), but the Pareto distribution has mainly been used to model (large) insurance losses since reinsurance premiums have simple and analytical expression, as in \citeN{Vajda:1951}, even for more complex treaties than standard stop-loss or excess-of-loss treaties as in \citeN{Kremer1984}. \citeN{BeirlantTeugels1992}, \citeN{McNeil:1997} (discussed in \shortciteN{Resnick:1997}) and \citeN{EKM} discussed more applications in insurance, with connections to extreme value theory.

The Value-at-Risk has been the finance benchmark risk measure for the past 20 years. Estimating a quantile for small probabilities, such as the 1\%-quantile of the profit and loss distribution for the next 10 days. Historically, practitioners used a standard Gaussian model, and then multiply by a factor of $3$ 
(as explained in \shortciteN{Klup:04}, this factor 3 is supposed to account for certain observed effects, also due to the model risk; it is based on backtesting procedures and can be increased by the regulatory authorities, if the backtesting proves the factor 3 to be insufficient). A less {\em ad hoc} strategy is to use a distribution that fits better tails of profit and loss distribution, such as the Pareto one.

Nevertheless, if the Pareto distribution remains extremely popular because several quantities can easily be derived (risk measures or insurance premium), in practice, distributions are only Pareto in very high tails (say above the 99.9\% quantile). So it becomes necessary to take into {account} second order approximation, to provide a better fit. Thus, we will see in this article how to derive risk measures and insurance premiums when distribution are Pareto-type above a (high) threshold $u$.

In Section \ref{sec:2}, we will define the two popular Pareto models, the strict Pareto distribution and the Generalized Pareto distribution (GPD), emphasizing differences between the two. A natural extension will be related to the concept of Pareto-type distribution, using regular variation function. In Section \ref{sec:3}, we will get back on the concepts of regular variation, and present the Extended Pareto distribution (EPD), introduced in \shortciteN{BeJoSe:09}. The inference of Pareto models will be discussed in Section \ref{sec:4}, with a discussion about the use of Hill estimator, still very popular in risk management, and Maximum Likelihood estimation. In Section \ref{sec:5} we will define classical measures and indices, and discuss expression and properties when underlying distribution are either strict Pareto, or Pareto-type, with high quantiles ($Q$), expected shortfall (ES), and large claim or top share indices (TS). And as Pareto models are usually assumed above a high threshold $u$ we will see, at the end of that section, how analytical expressions of those measures can be derived when Pareto-type distributions are considered only above threshold $u$ (and define $Q_u$, $\text{ES}_u$, etc). Finally, applications on real data will be considered, with insurance and reinsurance pricing in Section \ref{sec:6} and log-returns and classical financial risk measures in Section \ref{sec:7}.



\section{Strict Pareto models}\label{sec:2}

Pareto models are obtained when the loss distribution $\mathbb{P}[X>x]$ exhibits a power decay $x^{-\alpha}$. From a mathematical perspective, such a power function is interesting because it is the general solution of Cauchy's functional equation (its multiplicative version), $h(x\cdot y)=h(x)\cdot h(y)$, $\forall x,y$. A probabilistic interpretation is the stability of the conditional distribution $X$ given $X>u$ for any threshold $u$: the conditional distribution still exhibits a power decay, with the same index. But if we try to formalize more, two distributions will be obtained: a Pareto I distribution will be obtained when modelling the distribution of {\em relative excesses} $X/u$ given $X>u$ - see Equation 
(\ref{eq:Pareto_I}) - while the GPD (Generalized Pareto Distribution) will be obtained when modelling the distribution of {\em absolute excesses} $X-u$ given $X>u$ - see Equation (\ref{eq:Pareto_II}).

\subsection{Pareto I distribution}

A Pareto Type I distribution, bounded from below by $u>0$, with tail parameter $\alpha$, has probability density function and cumulative density function (CDF) equal to
\begin{equation}
f(x) = \frac{\alpha u^\alpha}{x^{\alpha+1}} \qquad\text{and}\qquad F(x) = 1-\left(\frac{x}{u}\right)^{-\alpha}, \qquad\text{for } x\geq u
\label{eq:Pareto_I}
\end{equation}
 If a random variable $X$ has distribution (\ref{eq:Pareto_I}), we will write $X\sim {\cal P}_1(u,\alpha)$. This distribution has an attractive property: the average above a threshold is proportional to the threshold, and it does not depend on the scale parameter $u$,
\begin{equation}
\mathbb{E}(X|X>u')=  \frac{\alpha u'}{\alpha-1}, \qquad \alpha>1
\label{eq:mean_over_u_ParetoI}
\end{equation}
where $u'\geq u$. Such a function is related to the {\em mean excess function} -- or {\em expected remaining lifetime} at age $u'$ when $X$ denotes the (random) life length -- defined as
\begin{equation}
e(u') = \mathbb{E}(X-u'|X>u').
\end{equation}
In the case of a Pareto distribution, it is also a linear function
\begin{equation}
e(u') = \frac{u'}{\alpha-1}, \qquad \alpha>1
\end{equation}

\begin{Remark}
In several textbooks and articles, Pareto models are defined with tail index $\xi=1/\alpha$, so that the survival function is proportional to the power function $x^{-1/\xi}$, instead of $x^{-\alpha}$. In that case, we have the confusing expressions 
\begin{equation}
\mathbb{E}(X|X>u')=\frac{u'}{1-\xi}\quad\text{ and }\quad e(u')=\frac{\xi u'}{1-\xi}, \qquad \xi<1.
\end{equation}
\end{Remark}

\subsection{Generalized Pareto distribution}

A Generalized Pareto Distribution (GPD), bounded from below by $u\geq0$, with scale parameter $\sigma$ and tail parameter $\alpha$, has cumulative density function (CDF)
\begin{equation}
F(x) = 1-\left[1+ \left(\frac{x-u}{\sigma}\right)\right]^{-\alpha} \qquad\text{for } x\geq u
\label{eq:Pareto_II}
\end{equation}
where $\sigma>0$ and $\alpha\in(0,\infty]$.  If a random variable $X$ has distribution (\ref{eq:Pareto_II}), we will write $X\sim {\cal GPD}(u,\sigma,\alpha)$.  The ${\cal GPD}(0,\sigma,\alpha)$ distribution is also called ``Lomax'' in the literature \shortciteN{Lomax:54}. The GPD is ``general'' in the sense that Pareto  I distribution is a special case, when $\sigma=u$:
\begin{equation}
{\cal{GPD}}(u,u,\alpha) = {\cal{P}}_1(u,\alpha)
\label{eq:Pareto_I_and_II}
\end{equation}

\citeN{Scollnik07} suggested an alternative expression, instead of (\ref{eq:Pareto_II}),
\begin{equation}
F(x) = 1-\left[\frac{u+\lambda}{x+\lambda}\right]^{\alpha} \qquad\text{for } x\geq u
\label{eq:Pareto_IIb}
\end{equation}
where $\lambda=\sigma-u$. Note that this function is used in \citeN{GAMLSS} to introduce a regression type model, where both $\lambda$ and $\alpha$ will be the exponential of linear combinations of some covariates, as in the \texttt{gamlss} package, in \texttt{R}.

\begin{Remark}
In several textbooks and articles, not only the Generalized Pareto has tail index $\xi=1/\alpha$, but $\xi$ appears also in the fraction term
\begin{equation}
F(x) = 1-\left[1+ \xi\left(\frac{x-u}{\sigma}\right)\right]^{-1/\xi} \qquad\text{for } x\geq u,
\label{eq:Pareto_II_bis}
\end{equation}
which could be seen as a simple alternative expression of the shape parameter $\sigma$. This expression is interesting since the limit when $\xi$ tends to $0$ can easily be defined, as well as the case where $\xi <0$, which could be interesting in extreme value theory.
\end{Remark}

The average above a higher threshold, $u' \geq u$, depends now on all parameters of the distribution, 
\begin{equation}
\mathbb{E}(X|X>u')=  \frac{\sigma-u}{\alpha-1} + \frac{\alpha }{\alpha-1} u',\
\label{eq:mean_over_u_ParetoII}
\end{equation}
The linearity of this function (as well as the mean excess function) characterizes the GPD class (see \shortciteN{GuessProschan:88} and \shortciteN{GhoshResnick:10}). A sketch of the proof is that we want
\begin{equation}
e(u')=\mathbb{E}(X|X>u')=\frac{1}{\overline{F}(u')}\int_{u'}^\infty xdF(x) = A+Bu',
\end{equation}
or, when differentiating with respect to $u'$, we obtain
\begin{equation}
B \overline{F}(u')du'=s -\big[A+(B-1)u'\big]d\overline{F}(u'),
\end{equation}
thus, up to some change of parameters
\begin{equation}
\frac{d\overline{F}}{\overline{F}} = -\alpha \frac{du}{u+c}\quad\text{ with }\quad\alpha=\frac{B}{B-1}
\end{equation}
which is a GPD distribution, with tail index $\alpha$.

One of the most important result in the extreme value theory states that, for most heavy-tailed distributions, the conditional excess distribution function, above a threshold $u$, converges towards a GPD distribution as $u$ goes to infinity (\shortciteN{Pick:75}, \shortciteN{BaHa:74}), for some parameters $\alpha$ and $\sigma$,
\begin{equation}
{F}_u(x) \longrightarrow \text{ GPD (or Pareto II) \quad as } u\rightarrow +\infty
\label{eq:theorem}
\end{equation}
where ${F}_u(x)=P(X-u\leq x | X>u)$. This result is known as the Pickands-Balkema-de Haan theorem, also called the second theorem in extreme value theory (as discussed in \cref{foot5}, \cpageref{foot5}). It provides strong theoretical support for modelling the upper tail of heavy-tailed distributions with GPD, {also known as the} Pareto Type II distribution. In a very general setting, it means that there are $\alpha$ and $\sigma$ such that $F_u$ can be approximated by the CDF of a $\mathcal{GPD}(u,\sigma,\alpha)$, see  \shortciteANP{EKM} (\citeyearNP{EKM}, Theorem 3.4.13, p.165).

\subsection{Threshold selection}

Whether to use a Pareto I or GPD model to fit the upper tail is related to the choice of the threshold. From (\ref{eq:Pareto_I}) and (\ref{eq:Pareto_II}), we have seen that Pareto I is a special case of GPD, when $\sigma=u$. They differ by an affine transformation when $\sigma\not=u$. A key property of Pareto distributions is that, if a distribution is Pareto for a fixed threshold $u$, it is also Pareto with the same tail parameter $\alpha$ for a higher threshold $u'\geq u$. For GPD, we have

\begin{equation}
\overline{F}_u={\cal{GPD}} (u,\sigma,\alpha)
\quad\Rightarrow\quad
\overline{F}_{u'}={\cal{GPD}}(u',\sigma+u'-u,\alpha).
\label{eq:trPareto2}
\end{equation}
where $\overline{F}_{u}$ is the survival excess function above $u$.\footnote{$\overline{F}_{u'}$ is a truncated Pareto distribution, with density equals to $f(x)/(1-F(u'))$. This property can be observed directly using Equation (\ref{eq:Pareto_IIb}), where both $\alpha$ and $\lambda$ remain unchanged.

Note that this property is quite intuitive, since the GPD distribution appears as a limit for exceeding distributions, and limit in asymptotic results are always fixed points: the Gaussian family is stable by addition (and appears in the Central Limit Theorem) while Fr\'echet distribution is max-stable (and appears in the first theorem in extreme value theory).} Thus, Pareto I and GPD are the same for all $u'\geq u$, if $\sigma=u$. Otherwise, we have $\sigma+u'-u\approx u'$ for very large values of $u'$ only. It follows that a GPD above a threshold will behave approximately as a Pareto I  above a {\em higher} threshold, much higher as $\sigma$ differs from $u$. This point will be discussed further in section \ref{sect:pq-indices}.


\section{Pareto-type models}\label{sec:3}

Since our interesting in risk management is usually motivated by the description of so-called {\em downside risks}, or (important) losses, as described in \shortciteN{Roy:1952}. Thus, we do not need that have a (strict) Pareto distribution for all $x$'s, in $\mathbb{P}[X>x]$, but possibly only when $x$'s are large enough. In this section, we will introduce some regular variation concepts, used to model that asymptotic property, and exhibit a distribution that is not strictly Pareto, by introducing a second order effect: the {\em Extended Pareto Distribution}.

\subsection{First- and second-order regular variation}\label{sect:second-order}

The tail index $\alpha$ is related to the max-domain of attraction of the underlying distribution, while parameter $\sigma$ is simply a scaling parameter.\footnote{\label{foot5}Historically, extremes were studied through block-maximum - yearly maximum, or maximum of a subgroup of observations. Following \citeN{FiTi:28}, up to some affine transformation, the limiting distribution of the maximum over $n$ i.i.d observations is either Weibull (observations with a bounded support), Gumbel (infinite support, but light tails, like the exponential distribution) or Fr\'echet (unbounded, with heavy tails, like Pareto distribution). \citeN{Pick:75} and \citeN{BaHa:74} obtained further that not only the only possible limiting conditional excess distribution is GPD, but also that the distribution of the maximum on subsamples (of same size) should be Fr\'echet distributed, with the same tail index $\gamma$, if $\gamma>0$. For instance in the U.S., if the distribution of maximum income per county is Fr\'echet with parameter $\gamma$ (and if county had identical sizes), then the conditional excess distribution function of incomes above a high threshold is a GPD distribution with the same tail index $\gamma$.}
The shape of the conditional excess cumulative distribution function is a power function (the Pareto distribution) if the threshold is large enough. Tails are then said to be Pareto-type, and can be described using so called {\em regularly varying} functions (see \shortciteNP{BiGoTe:87}). 

First and second order regular variation were  originally used  in  extreme  value theory, to study respectively the tail behavior of a distribution and the speed of convergence of the extreme value condition (see \shortciteNP{BiGoTe:87}, \shortciteNP{HaSt:96}, \shortciteNP{PeQi:04}, or section 2 in \shortciteNP{HaFe:06} for a complete survey). 
A function $H$ is said to be regularly varying (at infinity) with index $\gamma\in\mathbb{R}$ if
\begin{equation}
\lim_{t\rightarrow\infty}\frac{H(tx)}{H(t)}=x^{\gamma}~\text{ or }~\lim_{t\rightarrow\infty}x^{-\gamma}\frac{H(tx)}{H(t)}=1.
\end{equation}
A function regularly varying with index $\gamma=0$ is said to be slowly varying. Observe that any regularly varying function of index $-\gamma$ can be written $H(x)=x^{-\gamma}\ell(x)$ where $\ell$ is some slowly varying function. 

Consider a random variable $X$, its distribution is regularly-varying with index $-\gamma$ if, up-to some affine transformation, its survival function is regularly varying. Hence,
\begin{equation}\label{eq:RV1:barF}
\lim_{t\rightarrow\infty}x^{-\gamma}\frac{\overline{F}(tx)}{\overline{F}(t)}=1
\end{equation} 
or
\begin{equation}
\overline{F}(x)= x^{-\gamma}\ell(x),   
\label{eq:RV}
\end{equation}
where $\overline{F}(x)=1-F(x)$. A regularly varying survival function is then a function that behaves like a power law function near infinity. Distributions with survival function as defined in (\ref{eq:RV}) are called {\em Pareto-type distributions}. It means that the survival function  tends to zero at polynomial (or power) speed as $x\rightarrow\infty$, that is, as $x^{-\gamma}$.
For instance, a Pareto I distribution, with survival function $\overline{F}(x)=x^{-\alpha} u^\alpha$, is regularly varying with index $-\alpha$, and the associated slowly varying function is the constant $u^\alpha$. And a GPD or Pareto II distribution, with survival function $
\overline{F}(x)=\left(1+\sigma^{-1}{x}\right)^{-\alpha}$,
is regularly varying with index $-\alpha$, for some slowly varying function. 
But in a general setting, if the distribution is not {\em strictly} Pareto, $\ell$ will not be constant, and it will impact the speed of convergence.

In \shortciteN{HaSt:96}, a concept of second-order regular variation function  is introduced, that can be used to derive a probabilistic property using the quantile function,\footnote{The quantile function $U$ is defined as $U(x)=F^{-1}(1-1/x)$.} as in \shortciteN{BeGoSeTe:04}. Following \shortciteN{BeJoSe:09}, we will consider distributions such that an extended version of equation (\ref{eq:RV1:barF}) is satisfied,
\begin{equation}
\lim_{t\rightarrow\infty}x^{-\gamma}\frac{\overline{F}(tx)}{\overline{F}(t)}=1+\frac{x^{\rho}-1}{\rho}, \text{ for some } \rho \leq 0,\label{eq:2nd-order-1}
\end{equation}
that we can write, up to some affine transformation, 
\begin{equation}
\overline{F}(x)=c x^{-\gamma}[1-x^{\rho}\ell(x)],   
\label{eq:2nd-order}
\end{equation}
for some slowly varying function $\ell$ and some second-order tail coefficient $\rho\leq 0$. The corresponding class of Pareto-type distributions defined in (\ref{eq:2nd-order}) is often named the Hall class of distributions, referring to \citeN{Hall:82}. It includes the Singh-Maddala (Burr), Student, Fr\' echet and Cauchy distributions. A mixture of two strict Pareto-I distributions will also belong to this class.

Since $\rho\leq 0$, $x\mapsto 1-x^\rho\ell(x)$ is slowly varying, and therefore, a distribution $\overline{F}$ that satisfies (\ref{eq:2nd-order}) also satisfies (\ref{eq:RV}). More specifically, in (\ref{eq:2nd-order}), the parameter $\rho$ captures the rate of convergence of the survival function to a strict Pareto distribution. Smaller is $\rho$, faster the upper tail behaves like a Pareto, as $x$ increases. 
Overall, we can see  that
\begin{itemize}
\item $\gamma$ is the first-order of the regular variation, it measures the  tail parameter of the Pareto distribution,  
\item $\rho$ is the second-order of the regular variation, it measures how much the upper tail deviates from a strictly Pareto distribution. 
\end{itemize}
In the following, we will write RV$(-\gamma,\rho)$. 
There are connections between tail properties of the survival function $\overline{F}$ and the density  $f$ (see also Karamata theory for first order regular variation). More specifically, if $\overline{F}$ is RV$(-\gamma,\rho)$, with $\gamma>1$, then $f$ is RV$(-\gamma-1,\rho)$.

For instance, consider a Singh-Maddala (Burr) distribution, with survival distribution $\overline{F}(x)=[1+x^a]^{-q}$, then a second order expansion yields
\begin{equation}
\overline{F}(x)=
x^{-aq}[1-q x^{-a}+o(x^{-a})]~\text{as}~x\rightarrow\infty
\label{eq:2nd-order-burr}
\end{equation}
which is regularly varying of order $- aq$ and with second order regular variation $-a$, that is RV$(-aq ,-a)$.

\subsection{Extended Pareto distribution}

\shortciteN{BeJoSe:09} show that  Equation (\ref{eq:2nd-order}) can be approximated by
\begin{equation}
 \overline{F}(x) = \left[x \left(1+\delta-\delta x^\tau\right)\right]^{-\alpha}  \qquad\text{for } x\geq 1
\label{eq:F_EPD}
\end{equation}
where $\tau\leq 0$ and $\delta > \max(-1,1/\tau)$.\footnote{Using the expansion $(1+y^a)^b \approx 1+b y^a$, for small $y^a$, in (\ref{eq:F_EPD}) yields  (\ref{eq:2nd-order}).}
The main feature of this function is that it  captures the second-order regular variation of the Hall class of distributions, that is, deviation to a strictly Pareto tail, as defined in \shortciteN{Hall:82},
\begin{equation}
 \overline{F}(x) = ax^{-\alpha}\left[1+ bx^{-\beta}+o(x^{-\beta})\right]
\end{equation}
or, using Taylor's expansion,
\begin{equation}
 \overline{F}(x) = ax^{-\alpha}\left[1+ b_1x^{-\alpha}+\cdots+b_kx^{-k\alpha}+o(x^{-k\alpha})\right]
\end{equation}
with general notations. For more details, see also \shortciteANP{AlBeTe:17} (\citeyearNP{AlBeTe:17}, section 4.2.1).

From (\ref{eq:F_EPD}), we can define the Extended Pareto Distribution (EPD), proposed by \shortciteN{BeJoSe:09}, as follows:
\begin{equation}
 F(x) = 1-\left[\frac{x}{u} \left(1+\delta-\delta \left(\frac{x}{u}\right)^\tau\right)\right]^{-\alpha}  \qquad\text{for } x\geq u
\label{eq:EPD}
\end{equation}
where $\tau\leq 0$ and $\delta > \max(-1,1/\tau)$. If a random variable $X$ has (\ref{eq:EPD}) as its CDF, we will write $X\sim {\cal EPD}(u,\delta,\tau,\alpha)$. 

Pareto I is a special case when $\delta=0$ and GPD is a special case when $\tau=-1$:
\begin{eqnarray}
&{\cal EPD}(u,0,\tau,\alpha) & = \quad{\cal P}_1 (u,\alpha) \\
&{\cal EPD}(u,\delta,-1,\alpha)& = \quad {\cal GPD} (1,u/(1+\delta),\alpha)
\end{eqnarray}

The mean over a threshold for the EPD distribution has no closed form expression.\footnote{ \shortciteANP{AlBeTe:17} (\citeyearNP{AlBeTe:17}, section 4.6) give an  approximation, based on $(1+\delta-\delta y^\tau)^{-\alpha}\approx 1-\alpha\delta+\alpha\delta y^\tau$, which can be very poor. Thus, we do not recommend to use it.} Numerical methods can be used to calculate it. Since $u'\geq u>0$, $X$ given $X>u'$ is a positive random variable and 
\begin{equation}
\mathbb{E}[X|X>u'] = \int_0^\infty \overline{F}_{u'}(x)dx
\end{equation}
where $\overline{F}_{u'}(x)=\mathbb{P}[X>x|X>u']$ for $x>u$, i.e.
\begin{equation}
\overline{F}_{u'}(x) = \frac{\overline{F}(x)}{\overline{F}(u')}\quad\text{ where }\quad\overline{F}\text{ is the s.d.f. of }X
\end{equation}
Thus
\begin{equation}
\mathbb{E}[X|X>u'] = u'+\frac{1}{\overline{F}(u')}\int_{u'}^\infty \overline{F}(x)dx
\label{eq:Ex}
\end{equation}
The integral in Equation (\ref{eq:Ex}) can be computed numerically. Since numerical integration over a finite segment could be easier, we can make a change of variable ($1/x$) to obtain an integral over a finite interval:
\begin{equation}
E_{u'} = \int_{u'}^\infty \overline{F}(x)dx =  \int_0^{1/u'} \frac{1}{x^2} \,\overline{F}\left( \frac{1}{x}\right)  dx 
\label{eq:Eup}
\end{equation}

The Extended Pareto distribution  has a stable tail property: if a distribution is EPD for a fixed threshold $u$, it is also EPD for a higher threshold $u'\geq u$,  with the same tail parameter $\alpha$. Indeed, deriving a truncated EPD distribution, we find
\begin{equation}
\overline{F}_u={\cal{EPD}} (u,\delta,\tau,\alpha) \quad\Rightarrow\quad
\overline{F}_{u'}={\cal{EPD}}(u',\delta',\tau, \alpha).
\label{eq:trEPD}
\end{equation}
where $\delta'=\delta(u'/u)^\tau/[1+\delta-\delta(u'/u)^\tau]$.
A plot of  estimates of the tail index $\alpha$ for several thresholds would then be useful. If the distribution is Extended Pareto, a stable horizontal straight line should be plotted. This plot is similar to Hill plot for Hill estimates of $\alpha$ from Pareto I distribution. It is expected to be more stable if the distribution is not strictly Pareto. Indeed, {\shortciteANP{EKM} (\citeyearNP{EKM}, p.194)  and \citeANP{Resn:07} (\citeyearNP{Resn:07}, p.87) illustrate that the Hill estimator can perform very poorly if the slowly varying function is not constant in (\ref{eq:RV}). It can lead to very volatile Hill plots, also known as Hill  {\em horror} plots. This will be discussed in section \ref{sect:hill}.}

\section{Inference based on Pareto type distributions}\label{sec:4}


In this section, we briefly recall two general techniques used to estimate the tail index, and additional parameters: Hill estimator and maximum likelihood techniques. For further details, see \shortciteN{EKM}, \shortciteN{HaFe:06} or \shortciteN{BeGoSeTe:04}, and references therein. 

\subsection{Hill's Estimator}\label{sect:hill}

In the case where the distribution above a threshold $u$ is supposed to be $\mathcal{P}(u,\alpha)$, Hill's estimator of $\alpha$ is
\begin{equation}
\widehat{\alpha}_u=\left(\frac{1}{n_u}\sum_{i=1}^{n_u}\log(x_{n+1-i:n})-\log(u)\right)^{-1}
\end{equation}
This estimator is the maximum likelihood estimator of $\alpha$.

From Theorem 6.4.6 in \shortciteN{EKM} and section 4.2 in \shortciteN{BeGoSeTe:04}, if we assume that the distribution of $X$ is Pareto-type, with index $\alpha$, $\overline{F}(x)=x^{-\alpha}\ell(x)$, if $u\rightarrow\infty$, $n\rightarrow\infty$ and $n_u/n\rightarrow0$, then  $\widehat{\alpha}_u$ is a (strongly) consistent estimator of $\alpha$ when observations are i.i.d., and further, under some (standard) technical assumptions,
\begin{equation}\label{eq:asympt-norm}
    \sqrt{n_u}\big(\widehat{\alpha}_u-\alpha\big)\overset{\mathcal{L}}{\rightarrow}\mathcal{N}(0,\alpha^2).
\end{equation}
This expression can be used to derive confidence interval for $\alpha$, but also any index mentioned in the previous section (quantile, expected shortfall, top shares, etc) using the $\Delta$-method. For instance
\begin{equation}\label{eq:asympt-norm}
    \frac{\sqrt{n_u}Q(1-p)}{\sqrt{1+(\log[n_u/n]-\log[p])^2}}\big(\widehat{Q}_u(1-p)-Q(1-p)\big)\overset{\mathcal{L}}{\rightarrow}\mathcal{N}(0,\alpha^{-2}),
\end{equation}
as shown in Section 4.6. of \shortciteN{BeGoSeTe:04}.

Nevertheless, as discussed in \shortciteN{EKM} and \shortciteN{BeGoSeTe:04}, this asymptotic normality is obtained under some appropriate choice of $u$ (as a function of $n$) and $\ell$ : $u$ should tend to infinity sufficiently slowly, otherwise, $\widehat{\alpha}_u$ could be a biased estimator, even asymptotically. More specifically, in section \ref{sect:second-order}, we introduced second order regular variation: in Equation (\ref{eq:2nd-order}), assume that $\ell(x)=-1$, so that
\begin{equation}
\overline{F}(x)=c x^{-\alpha}[1+x^{\rho}]  
\end{equation}
From Theorem 6.4.9 in \shortciteN{EKM}, if $n_u=o(n^{(2\rho)/(2\rho+\alpha)})$, then the asymptotic convergence of Equation (\ref{eq:asympt-norm}) is valid. But if $n^{-(2\rho)/(2\rho+\alpha)}n_u$ tends to $\lambda$ as $n\rightarrow\infty$, then
\begin{equation}\label{eq:asympt-norm-biais}
    \sqrt{n_u}\big(\widehat{\alpha}_u-\alpha\big)\overset{\mathcal{L}}{\rightarrow}\mathcal{N}\left(\frac{\alpha^3\lambda}{\rho-\alpha},\alpha^2\right),
\end{equation}
has an asymptotic bias. Such a property makes Hill estimator dangerous to use for Pareto-type distributions.

\subsection{Maximum Likelihood Estimator}

For the GPD, a popular is to use the maximum likelihood estimator. When threshold $u$ is given, the idea is to fit a $\mathcal{GPD}(0,\sigma,\alpha)$ distribution on the sample $\{y_1,\cdots, y_{n_u}\}=\{x_{n+1-n_u:n}-u,\cdots, x_{n:n}-u\}$. The density being
\begin{equation}
f(y;\sigma,\alpha)=\frac{\alpha}{\sigma}\left(1+\frac{y}{\sigma}\right)^{-\alpha-1}
\end{equation}
the log-likelihood is here
\begin{equation}
\log\mathcal{L}(\sigma,\alpha;\boldsymbol{y})=-n\log\frac{\alpha}{\sigma}-(\alpha+1)\sum_{i=1}^{n_u}\left(1+\frac{y_i}{\sigma}\right).
\end{equation}
As expected, the maximum likelihood estimator of $(\sigma,\alpha)$ is asymptotically Gaussian, as shown in Section 6.5 of \shortciteN{EKM} (with a different parameterization of the GPD). In the case of the Extended Pareto distribution, there might be numerical complications, but \cite{BeJoSe:09} provided theoretical properties, and \shortciteN{ReIns} provided \texttt{R} codes, used in \shortciteN{AlBeTe:17}.

Note that since $\alpha$ is usually seen as the most important parameter (and $\sigma$ is more a nuisance parameter) it can be interesting to use profile likelihood techniques to derive some confidence interval, as discussed in Section 4.5.2 in \shortciteN{davison:2004}. Consider some Pareto type model, with parameter $(\alpha,\boldsymbol{\theta})$. Let 
\begin{equation}
    (\widehat{\alpha},\widehat{\boldsymbol{\theta}}) = \underset{\alpha,\boldsymbol{\theta}} {\text{argmax}}\left\lbrace\log\mathcal{L}(\alpha,\boldsymbol{\theta})\right\rbrace 
\end{equation}
denote the maximum likelihood estimator. From the likelihood ratio test, under technical assumptions,
\begin{align*}
2\big(\log\mathcal{L}(\widehat{\alpha},\widehat{\boldsymbol{\theta}})-\log\mathcal{L}(\alpha,\boldsymbol{\theta})\big) 
&\rightarrow \chi^2\big(\text{dim}(\alpha,\boldsymbol{\theta})\big)
\end{align*}
where $I_n(\alpha,\boldsymbol{\theta})$ is Fisher information. The idea of the profile likelihood estimator is to define, given $\alpha$
\begin{equation}
    \widehat{\boldsymbol{\theta}}_{\alpha} = \underset{\boldsymbol{\theta}} {\text{argmax}}\left\lbrace\log\mathcal{L}(\alpha,\boldsymbol{\theta})\right\rbrace 
\end{equation}
and then
\begin{equation}
    \widehat{\alpha}_{p} = \underset{\alpha} {\text{argmax}}\left\lbrace\log\mathcal{L}(\alpha,\boldsymbol{\theta}_\alpha)\right\rbrace 
\end{equation}
Then
\begin{align*}
2\big(\log\mathcal{L}(\widehat{\alpha}_p,\widehat{\boldsymbol{\theta}}_{\widehat{\alpha}_p})-\log\mathcal{L}(\alpha,\boldsymbol{\theta})\big) 
&\rightarrow \chi^2\big(1)\big).
\end{align*}
Thus, if $\log\mathcal{L}_p(\alpha)$ denotes the profile likelihood, defined as $\log\mathcal{L}_p(\alpha)=\log\mathcal{L}(\alpha,\widehat{\boldsymbol{\theta}}_{\alpha} )$, then a $95\%$ confidence interval can be obtained,
\begin{equation}
\big\lbrace\alpha: \log\mathcal{L}_p(\alpha) \geq \log\mathcal{L}_p(\widehat{\alpha}_p)-\frac{1}{2}q_{\chi^2(1)}(95\%)\big\rbrace
\end{equation}
where $q_{\chi^2(1)}(95\%)$ is the $95\%$ quantile of the $\chi^2(1)$ distribution.

\subsection{Application on Simulated Data}\label{sec:4:3}

\begin{figure}
    \centering
    \includegraphics[width=\textwidth]{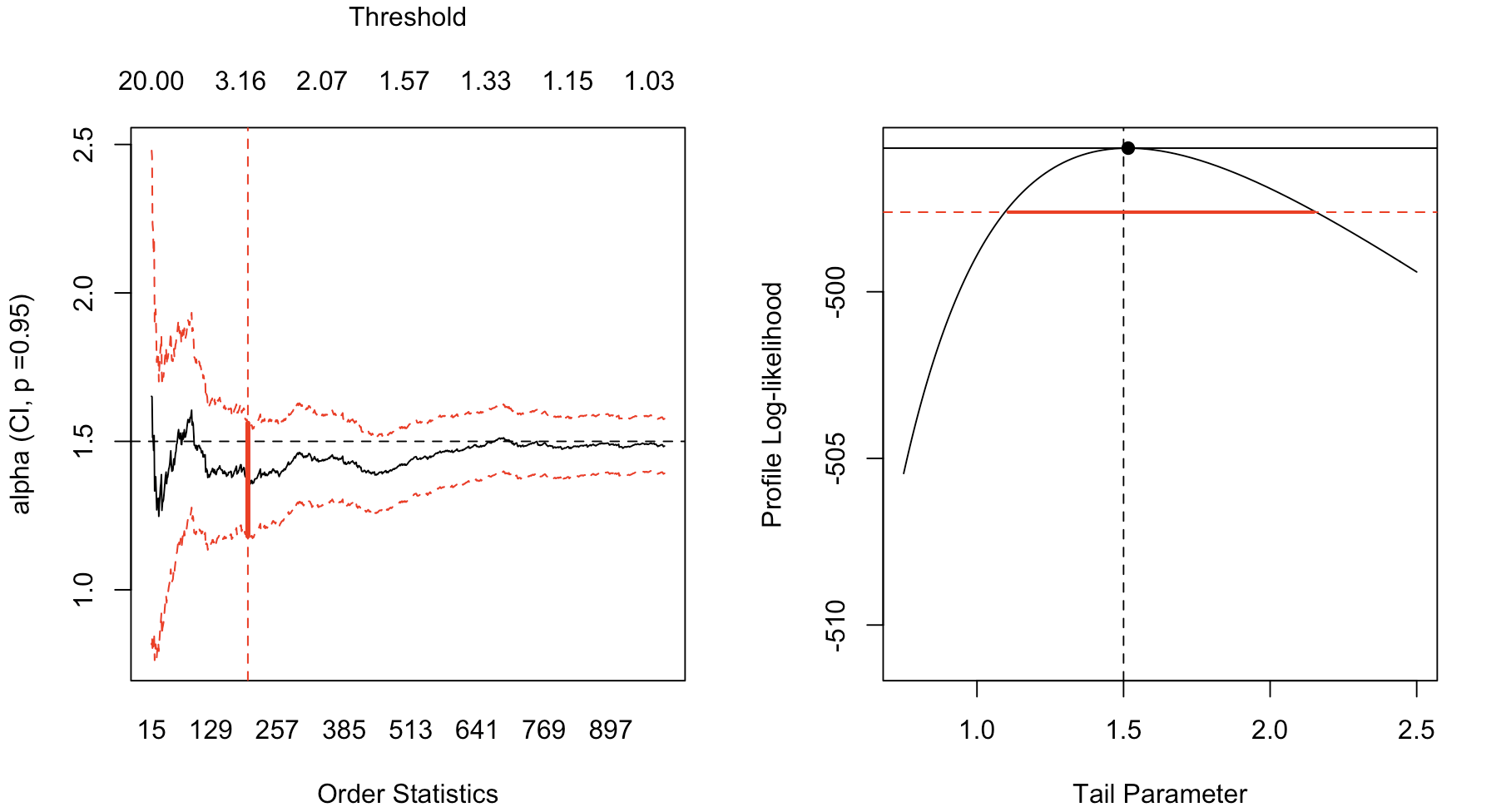}
    \caption{Hill estimator (on the left) for $\alpha$ and profile likelihood for $\alpha$ from a GPD model (on the right) on the top $20\%$ of the observations, when $n=1,000$ observations where generated from a (strict) Pareto distribution, with $\alpha=1.5$.}
    \label{fig:hill-PL:pareto1}
\end{figure}

\begin{figure}
    \centering
    \includegraphics[width=\textwidth]{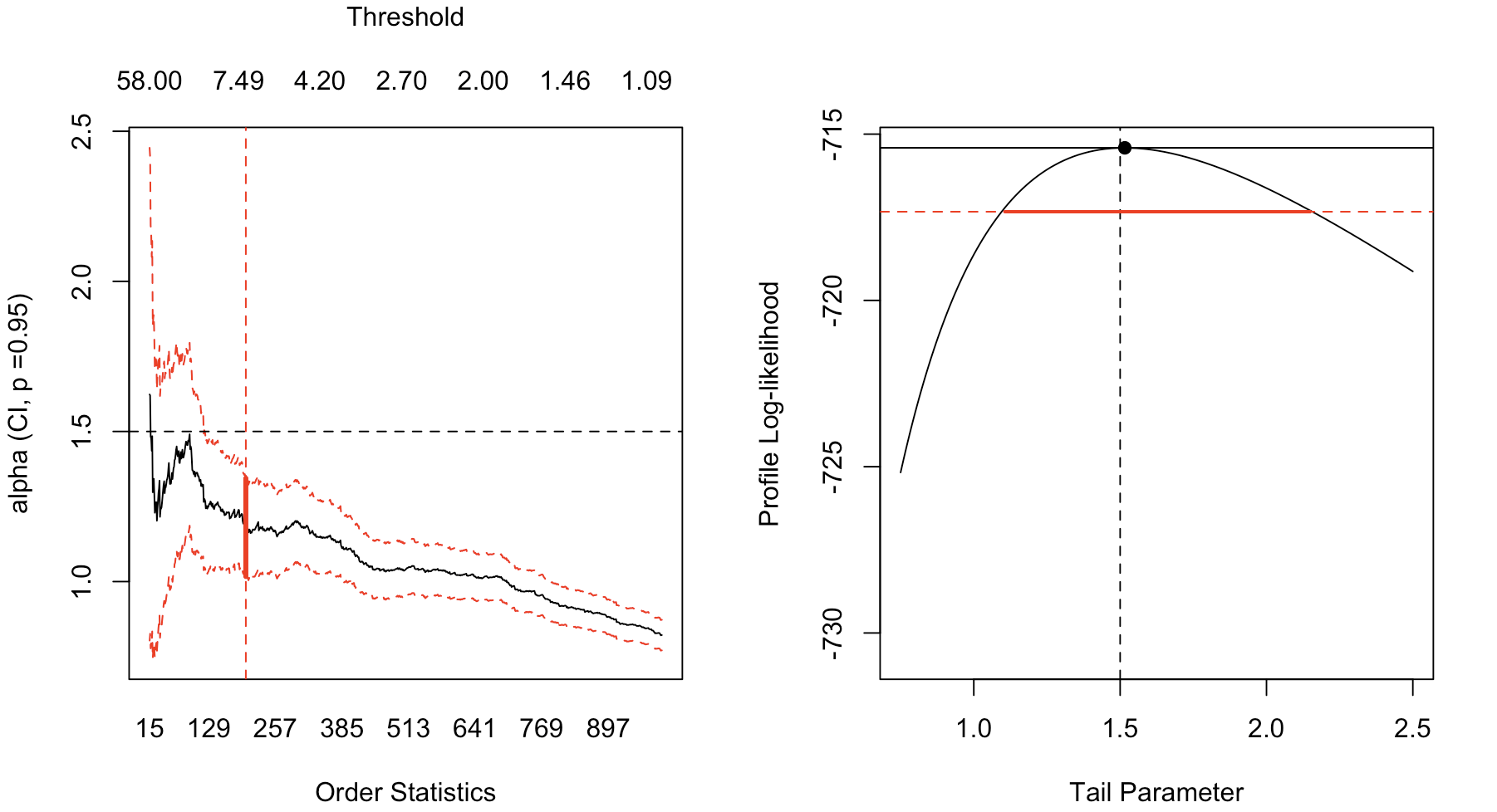}
    \caption{Hill estimator (on the left) for $\alpha$ and profile likelihood for $\alpha$ from a GPD model (on the right) on the top $20\%$ of the observations, when $n=1,000$ observations where generated from a Generalized Pareto distribution, with $\alpha=1.5$, and $\sigma=1$.}
    \label{fig:hill-PL:gpd}
\end{figure}

\begin{figure}
    \centering
    \includegraphics[width=\textwidth]{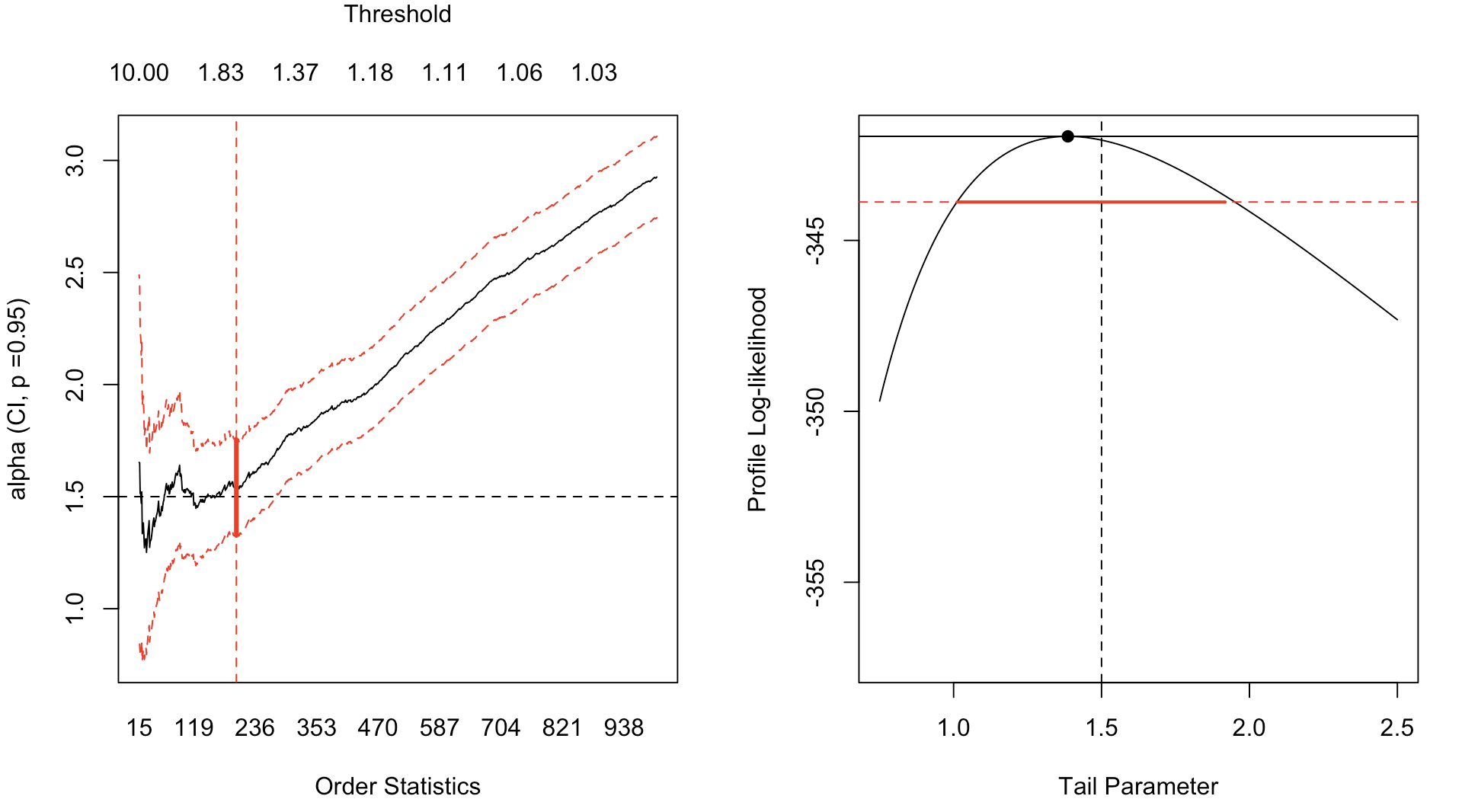}
    \caption{Hill estimator (on the left) for $\alpha$ and profile likelihood for $\alpha$ from a GPD model (on the right) on the top $20\%$ of the observations, when $n=1,000$ observations where generated from an extended Pareto distribution, with $\alpha=1.5$, and $\tau=-2$.}
    \label{fig:hill-PL:epd}
\end{figure}

In this section, three distributions where simulated: a strict Pareto on Figure \ref{fig:hill-PL:pareto1} with tail index $\alpha=1.5$, a Generalized Pareto on Figure \ref{fig:hill-PL:gpd} with tail index $\alpha=1.5$, and an Extented Pareto on Figure \ref{fig:hill-PL:epd} with tail index $\alpha=1.5$. Each time, on the left, the Hill plot is plotted, i.e. the plot $u\mapsto \widehat{\alpha}_u$. The dotted lines are bounds of the confidence interval (at level 95\%). The vertical segment (\textcolor{red}{\fulllwd}) is the confidence interval for $\alpha$ when $u$ is the $80\%$-quantile of the sample. On the right, the profile likelihood of a GPD distribution is plotted, including the horizontal line (\textcolor{red}{\fulllwd}) that defines the $95\%$-confidence interval.

\begin{figure}
    \centering
    \includegraphics[width=.48\textwidth]{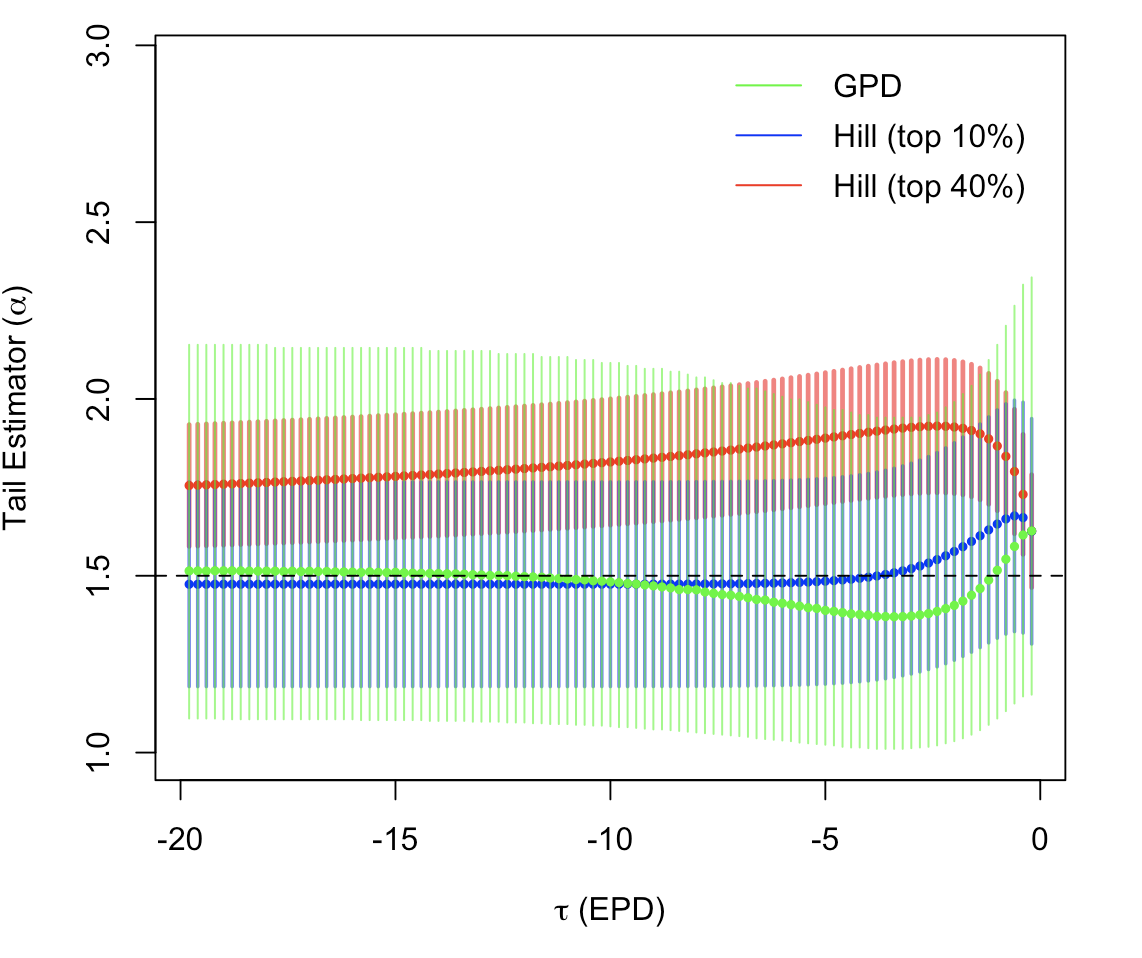}\includegraphics[width=.48\textwidth]{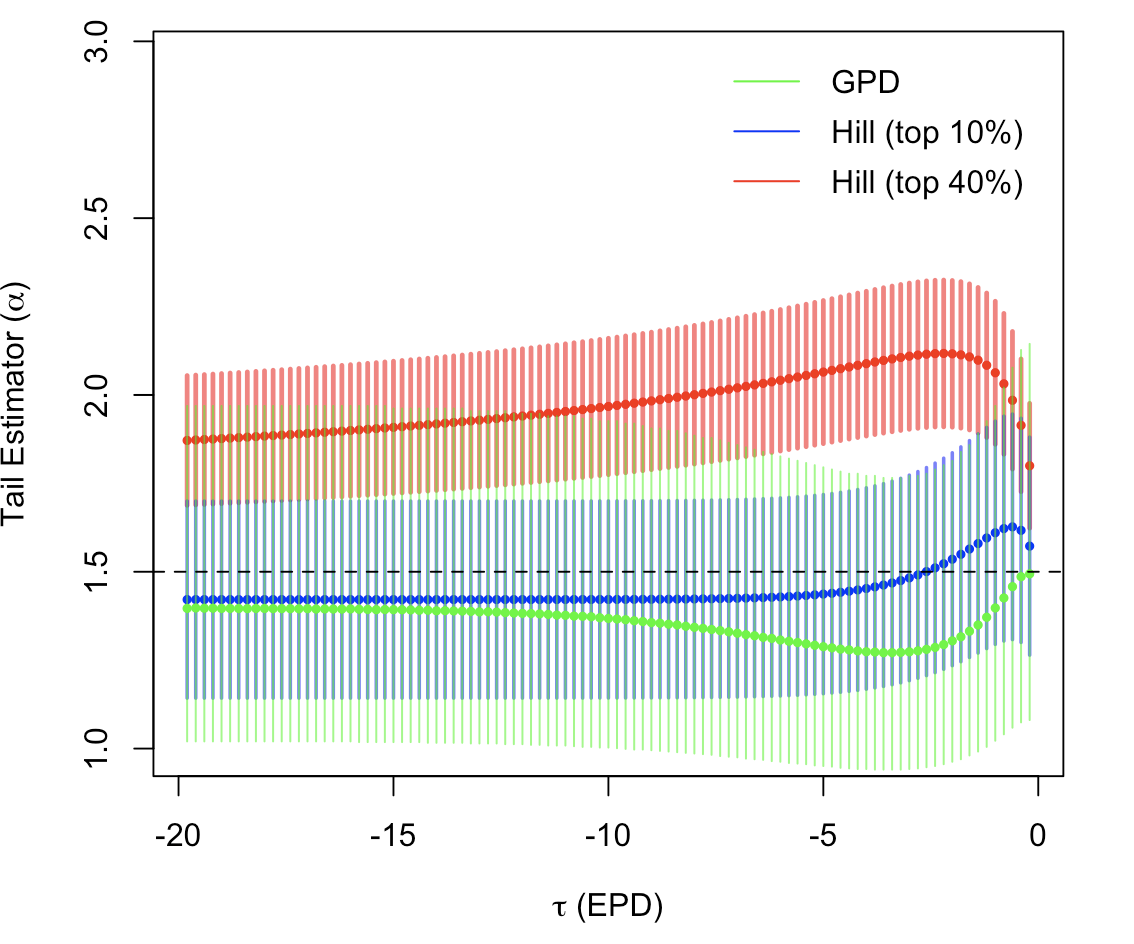}
    \caption{Simulation of EPD samples (with the same monte-carlo seed on the left, and another seed on the right), with $\alpha=1.5$, and $\tau$ varying from $-20$ to $0$. The red curve (\textcolor{red}{\fulllwd}) is Hill estimator $\widehat{\alpha}_u$ obtained when $u$ is the $60$\% quantile, while the blue curve (\textcolor{blue}{\fulllwd}) is obtained when $u$ is the 90\% quantile. The green curve (\textcolor{green}{\fulllwd}) is the profile likelihood estimator of a GPD distribution.}
    \label{fig:hill-PL-tau}
\end{figure}

In the case of the EPD, the smaller the value of $\tau$, the more bias Hill estimator has (see \citeN{CharpentierFlachaire} for a detailed discussion). When generating a Pareto type distribution, using the top 80\% observations does not guarantee anymore that Hill estimator could be a relevant estimator. On Figure \ref{fig:hill-PL-tau}, we used simulated samples of size $n=1,000$ with the same Monte Carlo seed (on the left), and another one (on the right), as on Figure \ref{fig:hill-PL:epd}, and we simply change the value of $\tau$. The red curve (\textcolor{red}{\fulllwd}) is Hill estimator $\widehat{\alpha}_u$ obtained when $u$ is the $60$\% quantile, while the blue curve (\textcolor{blue}{\fulllwd}) is obtained when $u$ is the $90$\% quantile. The green curve (\textcolor{green}{\fulllwd}) is the profile likelihood estimator of a GPD distribution. Horizontal segments are confidence intervals. For Hill estimator, observe that if the threshold $u$ is not large enough, estimator can be substantially biased. For the GPD, the fit is better, but the confidence interval is very large. Observe further that when $\tau\in(-2,0)$, Hill estimator in the higher tail can actually be overestimated.


\section{Modeling large events}\label{sec:5}

Given $n$ i.i.d. random losses $X_1,\cdots,X_n$, let $S_n$ and $M_n$ respectively denote the sum and the maximum
\begin{equation}
S_n=X_1+\cdots+X_n \quad\text{ and }\quad M_n=\max\lbrace X_1,\cdots,X_n\rbrace.
\end{equation}
A classical concept is the {\em Probable Maximum Loss} (section \ref{sub:sec:PML}) which is the distribution of $M_n$, or an affine transformation (since the distribution depends on $n$, the number of claims in insurance, or the time horizon in finance). It is possible, as in Section 3 of \citeN{BeGoSeTe:04}, to look ``{\em close to the maximum}'', by studying the limiting distribution of the $k$-th largest value\footnote{Given a sample $\lbrace x_1,\cdots,x_n \rbrace$, let $\lbrace x_{1:n},\cdots,x_{n:n} \rbrace$ denote the ordered version, with $x_{1:n}=\min\lbrace x_1,\cdots,x_n\rbrace$, $x_{n:n}=\max\lbrace x_1,\cdots,x_n\rbrace$ and $x_{1:n}\leq\cdots x_{n-1:n}\leq x_{n:n} $.} $X_{k:n}$, when $n$ goes to infinity, as well as $k$. If $k/n=O(1)$, we study the asymptotic distribution of some high quantile (Section \ref{sub:sec:HQ}).
It is also possible to focus on large losses with respect to the entire portfolio, i.e. $S_n$. This will yield the concepts of {\em Sub-Exponential} distribution (section \ref{sub:sec:SubEx}) when comparing $M_n$ and $S_n$, and comparing a high quantile, or the sum above a high quantile will be related to {\em Top Share} (section \ref{sub:sec:TS}), also called {\em Large Claim Index} in insurance applications.


\subsection{Probable Maximum Loss}\label{sub:sec:PML}

Paul Levy extended the 
Central Limit Theorem to variables with a non-finite variance by considering non-degenerate distributions $G$ such that
\begin{equation}
\lim_{n\rightarrow\infty}\mathbb{P}\left(\frac{\overline{X}_n-b_n}{a_n}\leq x\right)=G(x)
\end{equation}
where $\overline{X}_n=n^{-1}S_n$. It is an extension of the Central Limit Theorem in the sense that, if $X_i$'s have a finite variance and if $b_n=\mathbb{E}[X]$ while $a_n=\sqrt{n^{-1}\text{Var}[X]}$, then $G$ is the $\mathcal{N}(0,1)$ distribution, as well as any sequences such that $b_n\rightarrow\mathbb{E}[X]$ and $\sqrt{n}~a_n\rightarrow\sqrt{\text{Var}[X]}$, as $n\rightarrow\infty$. In the case of variables with infinite variance, then $G$ is called a stable distribution.

At the same period, Fisher-Tippett investigated a rather similar problem, searching for possible limiting distribution for a standardized version of the maximum
\begin{equation}
\lim_{n\rightarrow\infty}\mathbb{P}\left(\frac{M_n-b_n}{a_n}\leq x\right)=G(x).
\end{equation}
It was shown - in \shortciteN{Gnedenko:43} - that the only possible limit was the so-called extreme value distribution
\begin{equation}
G_{\alpha}(x)=\exp(-x^{-\alpha})
\end{equation}
including the limiting case $G_0(x)=\exp(-e^{-x})$, on some appropriate interval. For instance, assume that $X$ is $\mathcal{P}({1,\alpha})$ distributed, let $b_n=0$ and $a_n=U(n)=n^{1/\alpha}$, then
\begin{align}
\lim_{n\rightarrow\infty}\mathbb{P}\left(\frac{M_n-b_n}{a_n}\leq x\right)&=\lim_{n\rightarrow\infty}\left[F(a_nx+b_n)\right]^n=\lim_{n\rightarrow\infty}\left[F(n^{1/\alpha}x)\right]^n \\
&=\lim_{n\rightarrow\infty}\left(1-\frac{x^{-\alpha}}{n}\right)^n=\exp(-x^{-\alpha})
\end{align}

More generally, assume that $X$ is Pareto type with tail index $\alpha$, and consider $b_n=0$ and $a_n\sim U(n)$ as $n\rightarrow\infty$. From Proposition 3.3.2 in \shortciteN{EKM},
\begin{equation}
\lim_{n\rightarrow\infty}\left[F(a_nx+b_n)\right]^n=\exp\left(\lim_{n\rightarrow\infty}-n\overline{F}(a_nx+b_n)\right)
\end{equation}
and since
\begin{equation}
n\overline{F}(a_nx) \sim \frac{\overline{F}(a_nx)}{\overline{F}(a_n)}\rightarrow x^{-\alpha}\text{ as }n\rightarrow\infty
\end{equation}
therefore, we can obtain
\begin{equation}
\lim_{n\rightarrow\infty}\mathbb{P}\left(\frac{M_n-b_n}{a_n}\leq x\right)=\exp(-x^{-\alpha})
\end{equation}
as previously. Such a distribution is know as the Fr\'echet distribution with index $\alpha$.


\subsection{High Quantiles and Expected Shorfall}\label{sub:sec:HQ}

Given a distribution $F$, the quantile of level $p$ of the distribution is
\begin{equation}
Q(p)=\inf\big\lbrace y\in\mathbb{R}:F(y)\geq p\big\rbrace.
\end{equation}
Since we focus on high quantiles, i.e. $p\rightarrow1$, a natural related function is
\begin{equation}
U(x) = Q\left(1-\frac{1}{x}\right),
\end{equation}
and to study properties of that function as $x\rightarrow\infty$. 

In the previous section, we studied limiting behavior of $a_n^{-1}(M_{n}-b_n)$ as $n\rightarrow\infty$. Let $\widehat{F}_n$ denote the empirical cumulative distribution of $\lbrace X_1,\cdots, X_n\rbrace$,
\begin{equation}
\widehat{F}_n(x)=\frac{1}{n}\sum_{i=1}^n\boldsymbol{1}(X_i \leq x).
\end{equation}
Observe that $M_{n}=\widehat{F}_n^{-1}(1-1/n)=\widehat{U}_n(n)$, so $M_{n}$ should be `close' to $U(n)$. So it could make sense to study 
\begin{equation}
\lim_{n\rightarrow\infty}\frac{U(nx)-U(n)}{a_n}
\end{equation}
\shortciteANP{BeJoSe:09} (\citeyearNP{BeJoSe:09}, section 3.3) {show} that the only possible limit is 
\begin{equation}
h(x)=\frac{x^\gamma-1}{\gamma}
\end{equation}
If $U$ is the quantile function of $\mathcal{P}({1,\alpha})$, $U(x)=x^{1/\alpha}$, with auxiliary function\footnote{The study of the limiting distribution of the maximum of a sample of size $n$ {made us introduce} a normalizing sequence $a_n$. Here, a continuous version is considered - with $U(t)$ instead of $U(n)$ - and the sequence $a_n$ becomes the auxiliary function $a(t)$.} $a(t)=\alpha^{-1}t^{1/\alpha}$, then 
\begin{equation}
\frac{U(tx)-U(t)}{a(t)}=\alpha(x^{1/\alpha}-1)
\end{equation}
for all $x$, i.e. $\gamma=\alpha^{-1}$.

Assume that $X$ is $\mathcal{P}({u,\alpha})$, then 
\begin{equation}
Q(p)=u\cdot (1-p)^{-1/\alpha}
\end{equation}
and if $X$ is $\mathcal{GPD}({u,\sigma,\alpha})$, then 
\begin{equation}
Q(p)=u+\sigma \big[(1-p)^{-1/\alpha}-1\big]
\end{equation} 
For extended Pareto distributions quantiles are computed numerically.\footnote{See the R packages \texttt{ReIns} or \texttt{TopIncomes}.}

Another important quantity is the average of the top $p\%$, called expected shortfall,
\begin{equation}
\text{ES}(p) = \mathbb{E}(X|X>Q(1-p)) = \frac{1}{p}\int_{1-p}^1 Q(u)du.
\end{equation}
This quantity is closely related to the mean excess function, since $\text{ES}(p)=Q(1-p)+e(Q(1-p))$, where $e$ is the mean excess function. 
Assume that $X$ is $\mathcal{P}({u,\alpha})$, then 
\begin{equation}
\text{ES}(p)=\frac{\alpha}{\alpha-1}u(1-p)^{-1/\alpha}=\frac{\alpha }{\alpha-1}Q(p),
\end{equation}
and if $X$ is $\mathcal{GPD}({u,\sigma,\alpha})$, then $X|X>Q(1-p)$ is $\displaystyle{\mathcal{GPD}\left({Q(1-p),\sigma+Q(1-p)+u,\alpha}\right)}$ from Equation (\ref{eq:trPareto2}), and therefore 
\begin{equation}
\text{ES}(p)=\frac{\alpha Q(1-p)+\sigma-u}{\alpha-1}.
\end{equation}
For the extended Pareto model, only numerical approximations can be obtained.

\begin{figure}[ht]
\centering\includegraphics[width=.48\textwidth]{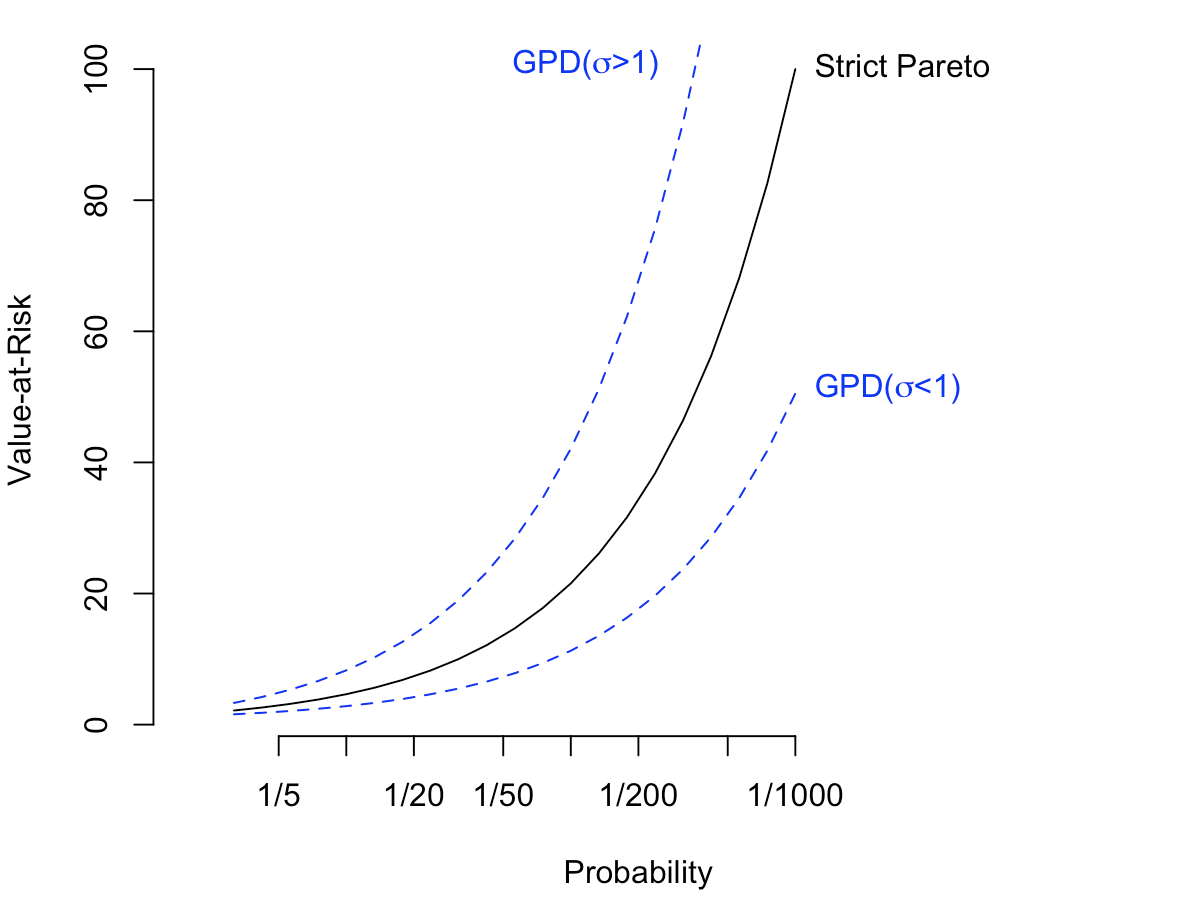}
\includegraphics[width=.48\textwidth]{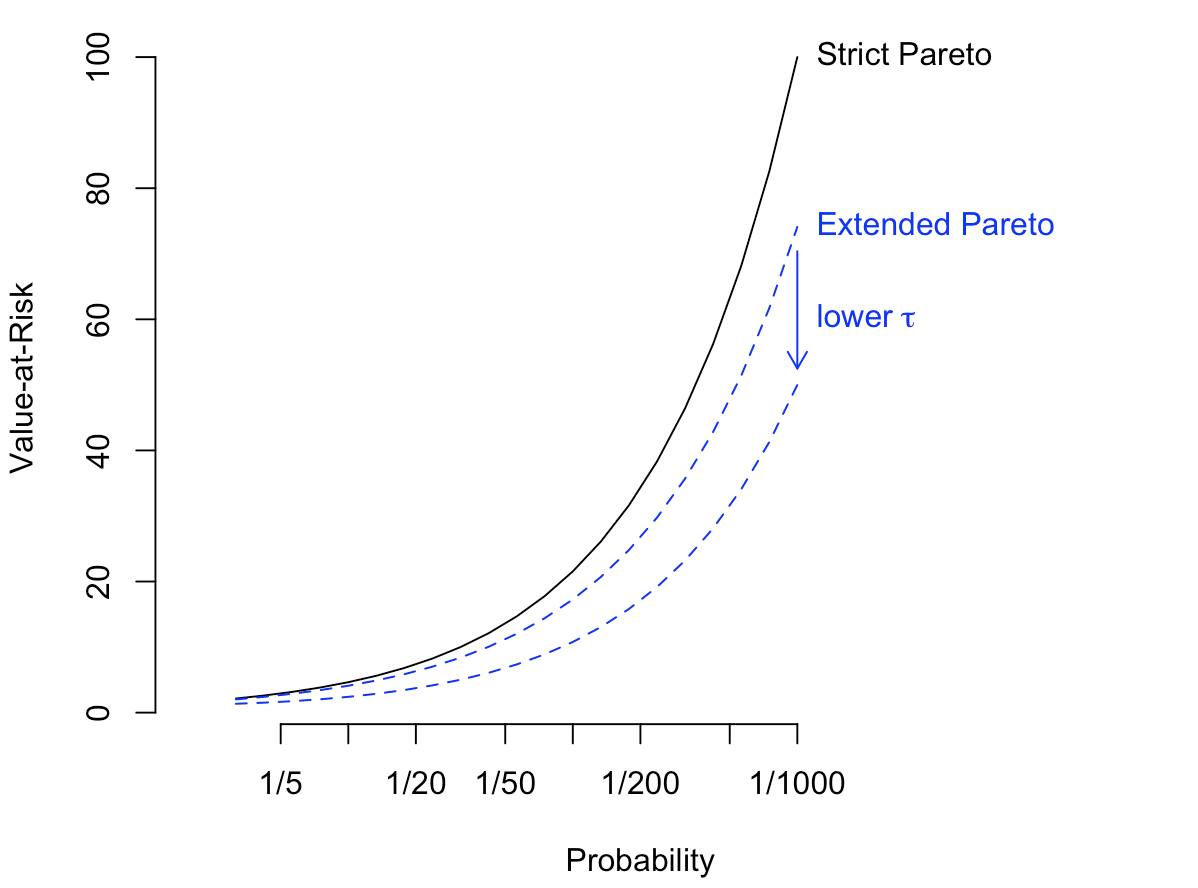}
\caption{Quantile $Q(1-p)$ as a function of the probability $p$ (on a log scale), with a strict Pareto (\full) with tail index $\alpha=1.5$, and a GPD distribution (\textcolor{blue}{\dashed}) on the left, and an EPD distribution (\textcolor{blue}{\dashed}) on the right.}\label{fig:VaR}
\end{figure}

On Figure \ref{fig:VaR} we have the evolution of $Q(1-p)$ for various values of $p$, from $1/10$ down to $1/1000$. A strict Pareto (\full) with tail index $\alpha=1.5$ is considered. On the left, we plot the quantile function of a GPD distribution (\textcolor{blue}{\dashed}), for small and large $\sigma$'s, and on the right, we consider an EPD distribution (\textcolor{blue}{\dashed}) with different values of $\tau$.

\begin{figure}[ht]
\centering\includegraphics[width=.48\textwidth]{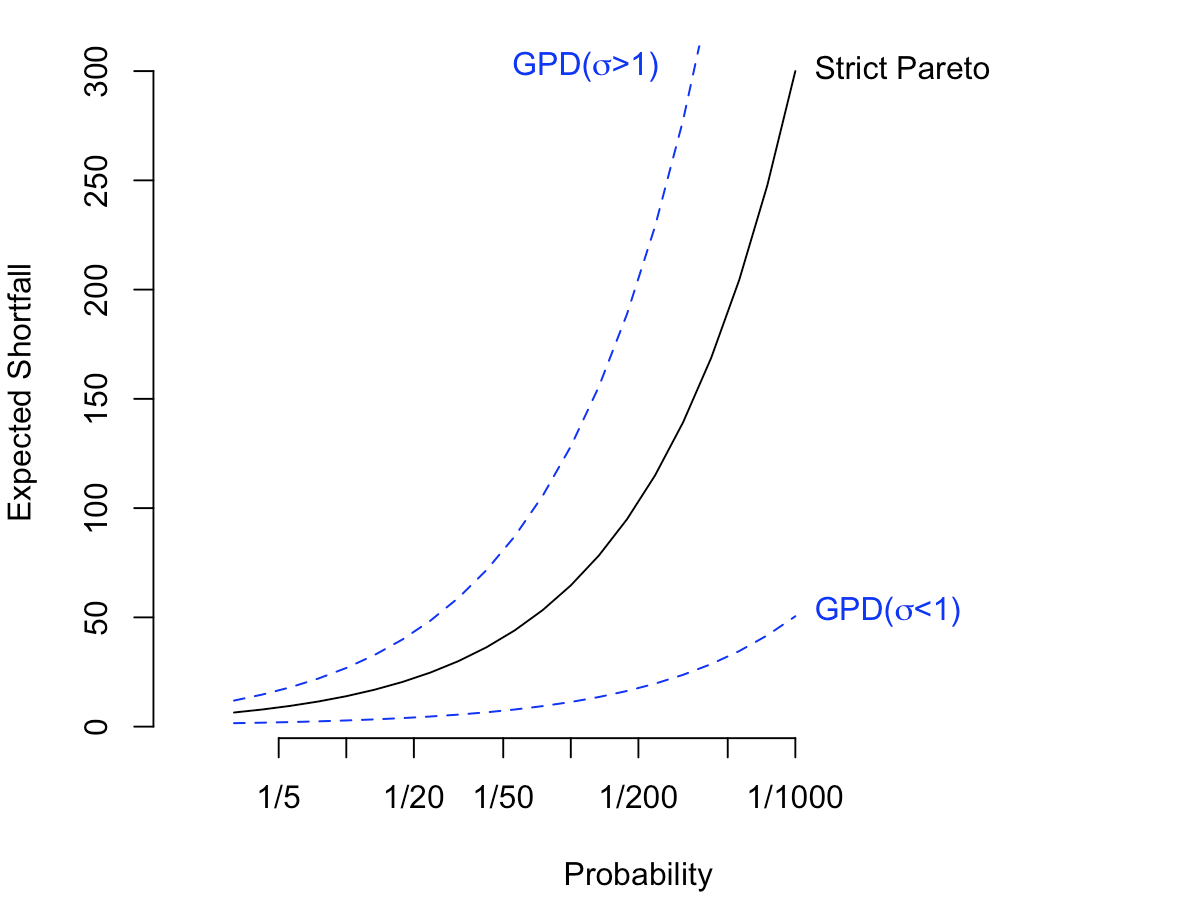}
\includegraphics[width=.48\textwidth]{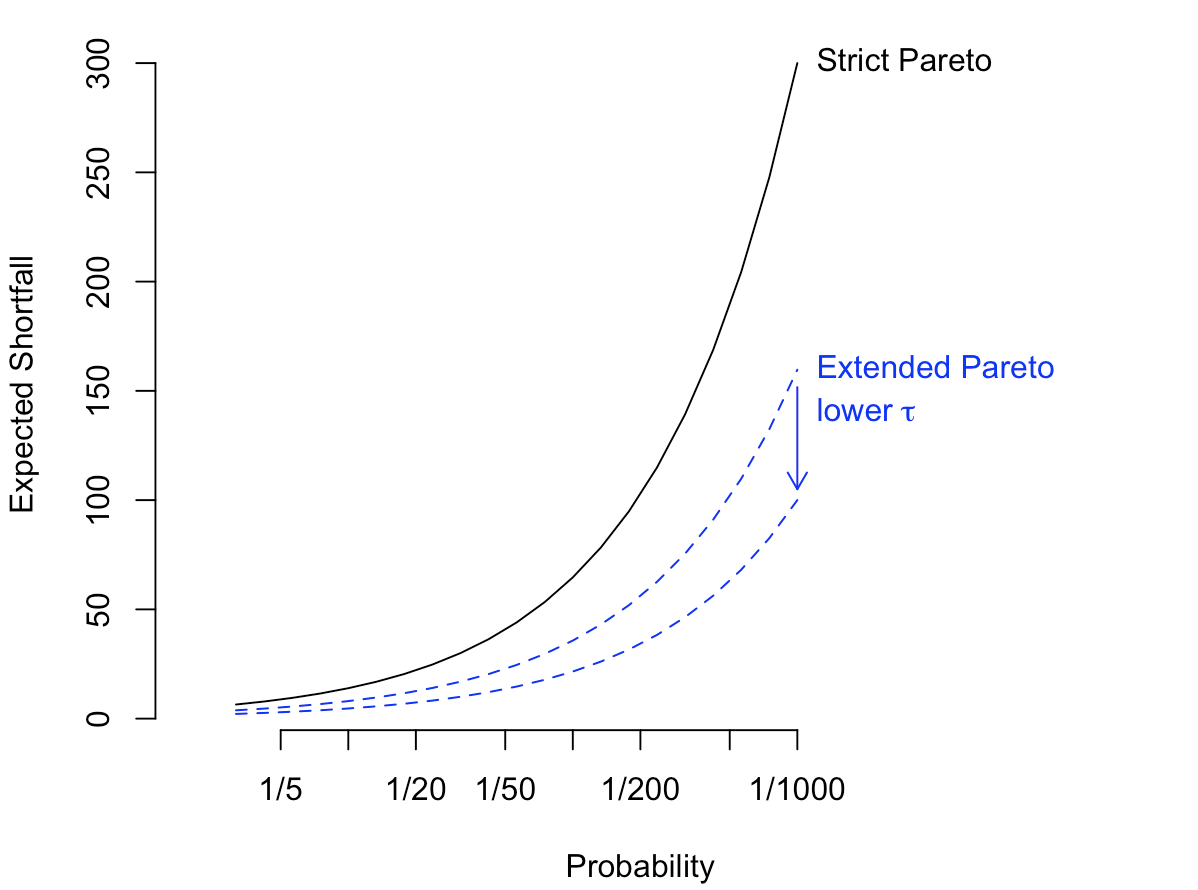}
\caption{Expected Shortfall $\text{ES}(p)$ as a function of the probability (on a log scale), with a strict Pareto (\full) with tail index $\alpha=1.5$, and a GPD distribution (\textcolor{blue}{\dashed}) on the left, and an EPD distribution (\textcolor{blue}{\dashed}) on the right.}\label{fig:ES}
\end{figure}

On Figure \ref{fig:ES} we have the evolution of $ES(p)$ for various values of $p$, from $1/10$ down to $1/1000$. The strict Pareto distribution (\full) still has tail index $\alpha=1.5$. On the left, we plot the expected shortfall of a GPD distribution (\textcolor{blue}{\dashed}), for small and large $\sigma$'s, and on the right, we consider an EPD distribution (\textcolor{blue}{\dashed}) with different values of $\tau$.

\begin{figure}[ht]
\centering\includegraphics[width=.48\textwidth]{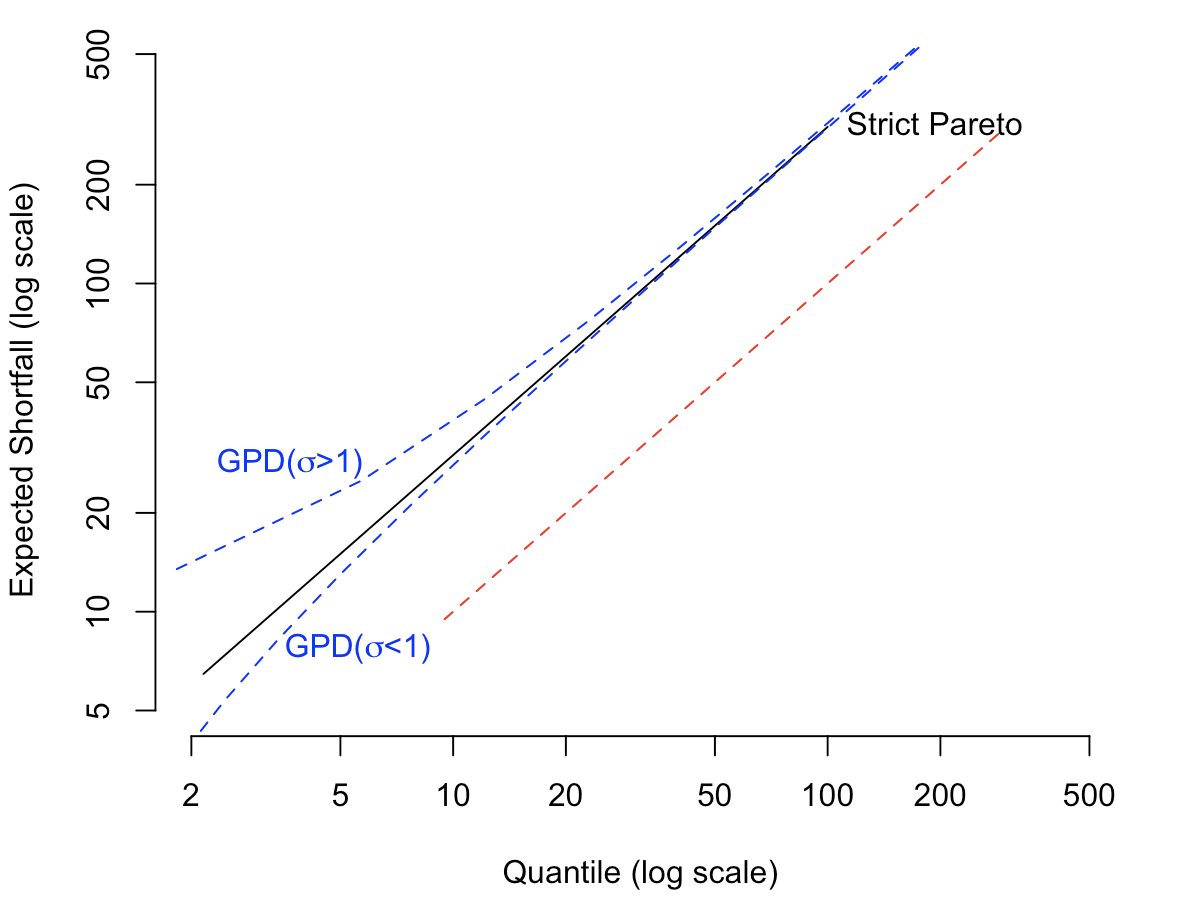}
\includegraphics[width=.48\textwidth]{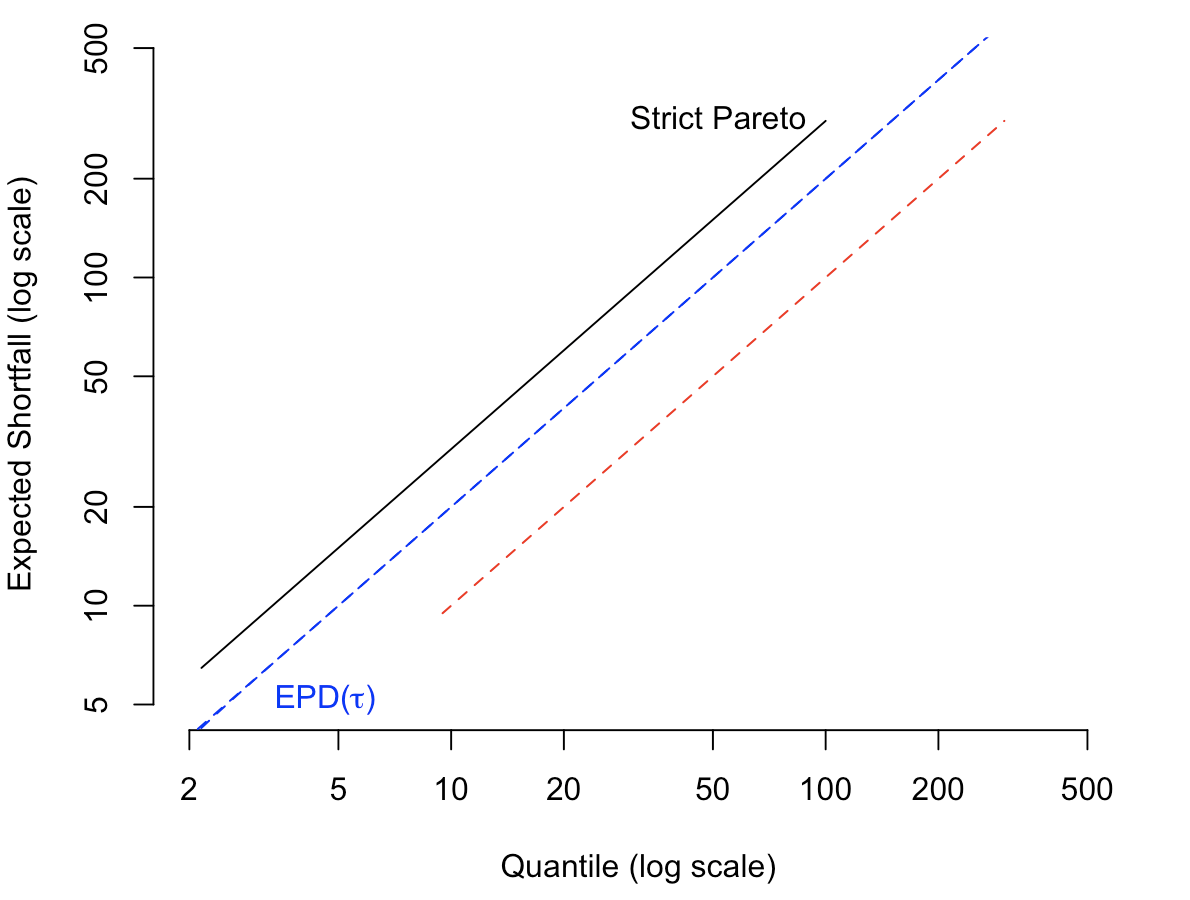}
\caption{Expected Shortfall against quantile for different values of the probability (i.e. $(Q(1-p)),\text{ES}(p)$), with a strict Pareto (\full) with tail index $\alpha=1.5$, and a GPD distribution (\textcolor{blue}{\dashed}) on the left, and an EPD distribution (\textcolor{blue}{\dashed}) on the right. The straight line below (\textcolor{red}{\dashed}) is the identity curve, where the Expected Shortfall would be equal to the associated quantile.}\label{fig:VaR-ES}
\end{figure}

Finally, on Figure \ref{fig:VaR-ES} we compare $Q(1-p)$ and $ES(p)$ for various values of $p$, and plot the line $(Q(1-p),ES(p))$. As expected, it is a straight line in the strict Pareto case (\full), above the first diagonal (\textcolor{red}{\dashed}), meaning that there can be substantial losses above any quantile. Two GPD distributions with the same tail index $\alpha$ are considered on the left (\textcolor{blue}{\dashed}), and an EPD case is on the right (\textcolor{blue}{\dashed}). The GPD case is not linear, but tends to be linear when $p$ is large enough (in the upper tail, we have a strict Pareto behavior). The EPD case is linear, with an Espected Shorfall smaller than the strict Pareto case.

\subsection{Maximum, Sum and Subexponential distributions}\label{sub:sec:SubEx}

In insurance, one possible way to define {\em large} losses is to say that a single claim is large when the total amount (over a given period of time) is predominantly determined by that single claim. The dominance of the largest value can be written
\begin{equation}
\mathbb{P}[S_n > x] \sim \mathbb{P}[M_n > x]\text{ as }x\rightarrow\infty.
\end{equation}
Heuristically, the interpretation is that the sum will exceed a large value $x$ only when (at least) one single loss $X_i$ also exceeds $x$. If losses $X_i$ are i.i.d. with CDF $F$ we will say that $F$ belongs to the class of subsexponential distributions, $F\in\mathcal{S}$. As proved in \shortciteANP{AlBeTe:17} (\citeyearNP{AlBeTe:17}, Section 3.2) strict Pareto and Pareto-type distributions belong to class $\mathcal{S}$. Hence, this class can be seen as an extension of the Pareto models. See \citeN{goldie98} for applications of such distributions in a risk-management context.

In sections 6.2.6 and 8.2.4 of \shortciteN{EKM}, another quantity of interest is introduced, the ratio of the maximum and the sum, $M_s/S_n$. As proved in \citeN{OBrien}, $\mathbb{E}(X)$ is finite if and only if the ratio converges (almost surely) towards 0. A non-degenerated limit {is obtained if and only if} $X$ is Pareto type, with index $\alpha\in(0,1)$. An alternative would be to consider the ratio of the sum of the largest claims (and not only the largest one), and the total sum.

\subsection{Top Share and Large Claim Index}\label{sub:sec:TS}

\begin{figure}[ht]
\centering\includegraphics[width=.48\textwidth]{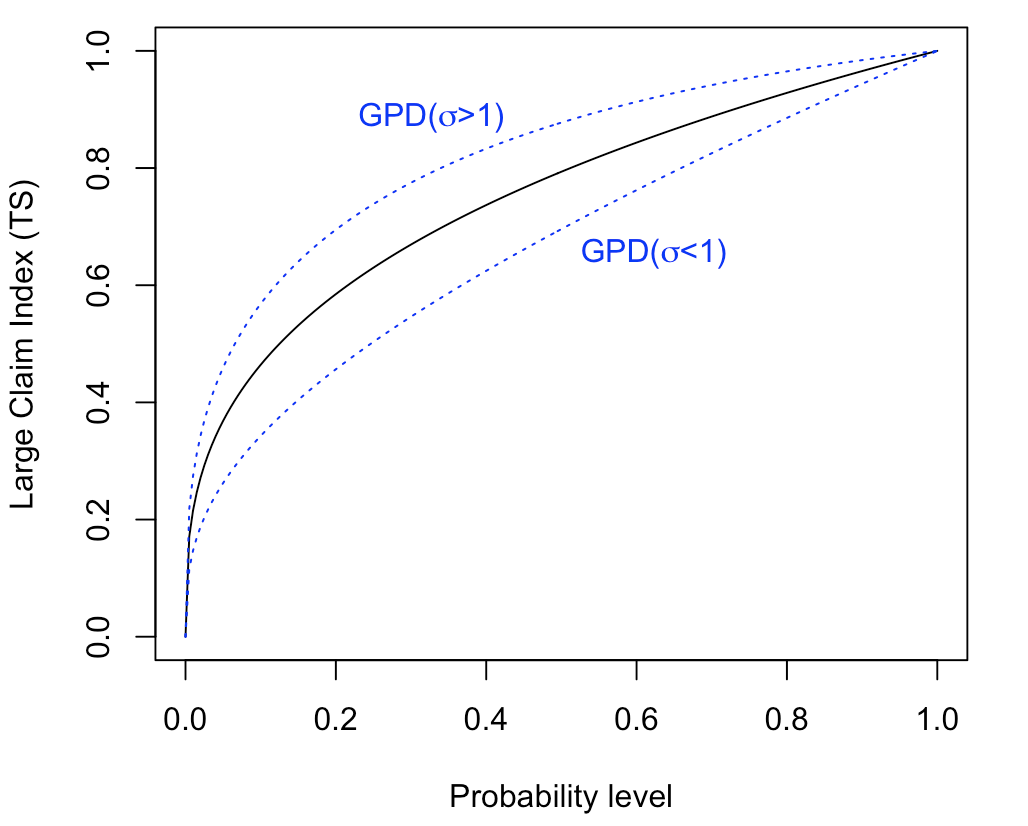}
\includegraphics[width=.48\textwidth]{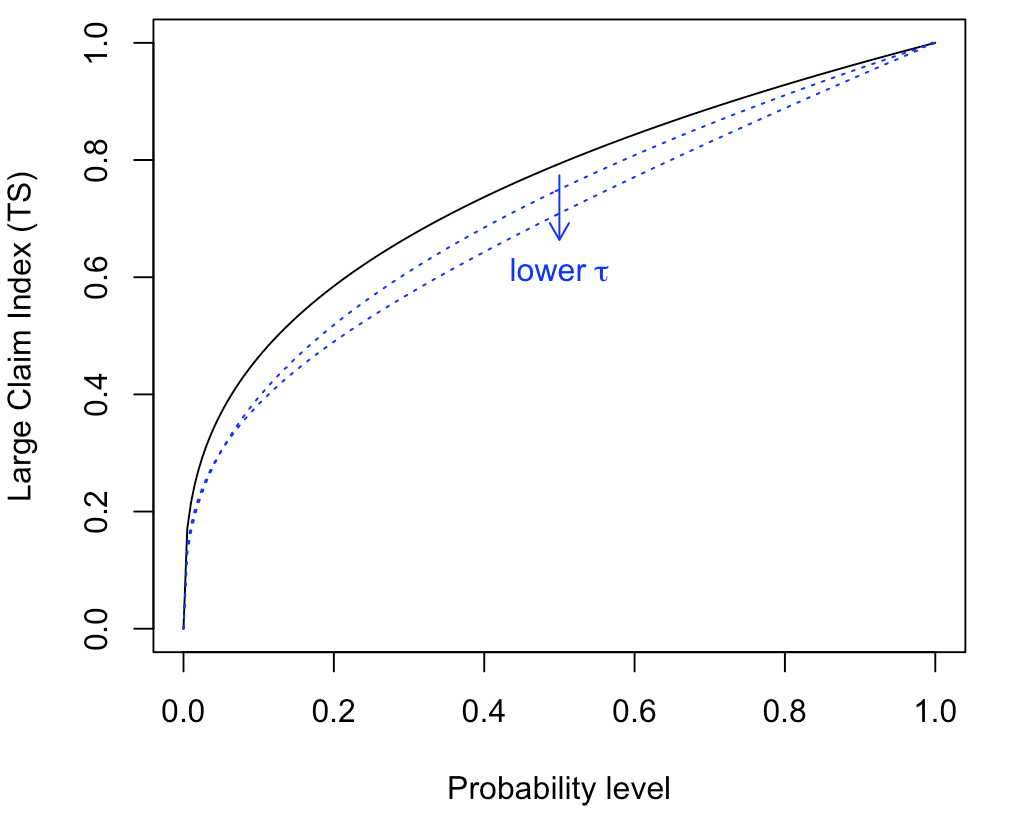}
\caption{Large Claim Index $\text{TS}(p)$ as a function of the probability $p$, with a strict Pareto (\full) with tail index $\alpha=1.5$, and a GPD distribution (\textcolor{blue}{\dashed}) on the left, and an EPD distribution (\textcolor{blue}{\dashed}) on the right.}\label{fig:LCI}
\end{figure}

In section \ref{sub:sec:HQ}, we discussed how to move from $M_n$ to the average of the top $p$\% of the distribution. It could be possible to normalize by the total sum. The top $p$-\% loss share can be  defined as 
\begin{equation}\label{eq:TS}
\text{TS}(p) = \frac{p\, \mathbb{E}(X|X>Q(1-p))}{\mathbb{E}(X)} =\frac{p\text{ES}(p)}{\mathbb{E}(X)}= 1-L(1-p)
\end{equation}
where $L$ denote the Lorenz curve, i.e.
\begin{equation}
L(u)=\frac{1}{\mathbb{E}(X)}\int_{0}^{Q(u)} \overline{F}(y)dy= \frac{1}{\mathbb{E}(X)}\int_{0}^{u} Q(y)dy.
\end{equation}
The empirical version of $\text{TS}_p$ is
\begin{equation}
\widehat{\text{TS}}(p)=\frac{x_{n-[np]:n}+x_{n:n}}{S_n}
\end{equation}
where $(x_{i:n})$ denote the order version of sample $(x_i)$, with $x_{1:n}\leq x_{2:n}\leq \cdots\leq x_{n:n}$. See section 8.2.2 of \shortciteN{EKM} for convergence 
properties of that estimator. In the case where $X$ is a strict Pareto $\mathcal{P}({u,\alpha})$, then 
\begin{equation}
\text{TS}(p)=p^{(\alpha-1)/\alpha},
\end{equation}
while if $X$ is $\mathcal{GPD}({u,\sigma,\alpha})$, 
\begin{equation}
\text{TS}(p)=1-\frac{u(1-p)+\alpha\cdot\big(I_{1}(\alpha-2,2)-I_{p^{1/\alpha}(\alpha-1,2)}\big)}{u+\sigma\alpha I_1(\alpha-1,2)}
\end{equation}
as in Section 6 of \shortciteN{Arno:08PA}, where $I_z(a,b)$ is the incomplete Beta function,
\begin{equation}
I_z(a,b)=\int_0^z t^{a-1}t^(1-t)^{b-1}dt.
\end{equation}
No simple expression can be derive for the extended Pareto case, but numerical computations are possible, as on Figure \ref{fig:LCI}, where strict Pareto distribution with tail index $\alpha=1.5$ is plotted against Pareto type distributions, GPD and EPD, with the same tail index.

\subsection{From Pareto to Pareto Type Models}\label{sect:pq-indices}

The concepts introduced earlier were based on the assumption that the distribution $F$ of the observations was known. But actually, for small enough $p$, we might only have a Pareto distribution in the upper tail, above some threshold $u$, with $\mathbb{P}[X> u]\geq p$, as suggested in \shortciteN{smith1987}. And let $q_u$ denote the probability to exceed that threshold $q_u$.

Note that if $x>u$, $\mathbb{P}[X> x]=\mathbb{P}[X> x|X>u]\cdot \mathbb{P}[X> u]=q_u \mathbb{P}[X> x|X>u]$. Thus, if a $\mathcal{P}(u,\alpha)$ distribution is considered in the tails,
\begin{equation}
\mathbb{P}[X> x]=q_u\left(\frac{x}{u}\right)^{-\alpha} 
\end{equation}
and therefore, 
\begin{equation}
Q_u(1-p) = u\cdot \left(\frac{p}{q_u}\right)^{-1/\alpha},\text{ where }q_u=\mathbb{P}[X> u],
\end{equation}
and
\begin{equation}
\text{ES}_u(p) = \frac{\alpha}{\alpha-1}Q_u(1-p) .
\end{equation}
If a $\mathcal{GPD}(u,\sigma,\alpha)$ distribution is considered in the tails,
\begin{equation}
Q_u(1-p) = u\cdot \sigma\left[\left(\frac{p}{q_u}\right)^{-1/\alpha}-1\right]^{-1},\text{ where }q_u=\mathbb{P}[X> u],
\end{equation}
and 
\begin{equation}
    \text{ES}_u(p)=\frac{\alpha Q_u(1-p)+\sigma-u}{\alpha-1}.
\end{equation}
Nevertheless, if a simple update of formulas was necessary for downside risk measures (such as a quantile and an expected shortfall), expression can be more complicated when they rely on the entire distribution, such as the Large Claim Index TS. From Equation (\ref{eq:TS}), if $p$ is small enough, write
\begin{equation}\label{eq:TS:2}
\text{TS}_u(p) = \frac{p\text{ES}_u(p)}{\mathbb{E}(X)}
\end{equation}
where $\mathbb{E}(X)$ is either approximated using $\overline{x}$, or a more robust version would be $(1-q_u)\overline{x}_u+q_u\text{ES}_u(q_u)$ where $\overline{x}_u$ is the empirical average of values below threshold $u$, while $\text{ES}_u(q_u)$ is the parametric mean from a Pareto-type model, about threshold $u$.


\section{Insurance and Reinsurance}\label{sec:6}

Classical quantities in a reinsurance context are the return period ({that can be related} to quantiles, and the financial Value-at-Risk) and a stop-loss premium (related to the expected-shortfall). 

\subsection{Return Period and Return Level}

Consider $Y_1,\cdots, Y_n$, 
a collection of i.i.d. random variables, with distribution $F$. Consider the sequence of i.i.d. Bernoulli variables $X_i=\boldsymbol{1}_{Y_i>z(t)}$ for some threshold $z(t)$, such that $\overline{F}(z(t))=1/t$. The first time threshold $z(t)$ 
is reached is the random variable $N_t$ defined as
\begin{equation}
N_t = \min\big\lbrace i\in\mathbb{N}_*:Y_i>z(t)\big\rbrace.
\end{equation}
Then $N_t$ has a geometric distribution, 
\begin{equation}
\mathbb{P}\big[N_t=k\big]=\frac{1}{(1-t)^{k-1}}\frac{1}{t}.
\end{equation}
The return period of the events $\{Y_i>z(t)\}$ is then $\mathbb{E}[N_t]=t$, or conversely, the threshold that is reached - on average - every $t$ events is $z(t)=Q(1-1/t)=U(t)$, called also return level.

Hence, with strict Pareto losses $\mathcal{P}(u,\alpha)$
\begin{equation}
z(t) = ut^{1/\alpha},
\end{equation}
while with $\mathcal{GPD}({u,\sigma,\alpha})$ losses,
\begin{equation}
z(t) = u+\sigma\left[t^{1/\alpha}-1\right].
\end{equation}
In the case of non-strict Pareto distribution, where only the top $q$\% is $\mathcal{P}(u,\alpha)$ distributed, 
\begin{equation}
z_u(t) = u(q_ut)^{1/\alpha},~\text{ where }~q_u=\mathbb{P}[X> u].
\end{equation}
Thus, if we assume that the top $10\%$ is Pareto distributed, the return level of a centennial event ($t=100$) corresponds to the return level of a decennial event ($q_ut=10$) for the top $10\%$ observations. If the top $q$\% is $\mathcal{GPD}(u,\sigma,\alpha)$ distributed, 
\begin{equation}
z_u(t) = u+\sigma\big((q_ut)^{1/\alpha}-1\big),~\text{ where }~q_u=\mathbb{P}[X> u].
\end{equation}
From an empirical perspective, consider a dataset with $n$ observations, i.i.d., $y_1,\cdots,y_n$. Assume that above threshold $u$, the (conditional) distribution is $\mathcal{GPD}({u,\sigma,\alpha})$. Denote $n_u$ the number of observations above $u$, so that $\widehat{F}_n(u)=1-n_u/n$. Then
\begin{equation}
\widehat{z}_u(t) = u+\widehat{\sigma}\left[\left(\frac{n_u}{n}t\right)^{1/\widehat{\alpha}}-1\right].
\end{equation}

\subsection{Reinsurance Pricing}

Let $Y$ denote the loss amount of an accident. Consider a contract with deductible $d$, so that the indemnity paid by the insurance company is $(Y-d)_+$, where $\cdot_+$ denotes the positive part, i.e. $(y-d)_+=\max\lbrace0,y-d\rbrace$.
The pure premium of such a contract, with deductible $d$, is 
\begin{equation}
\pi(d)=\mathbb{E}\big[(Y-d)_+\big]
\end{equation}
and therefore, if $e(d)$ denotes the mean excess function,
\begin{equation}
\pi(d)=\mathbb{E}\big[Y-d|Y>d\big]\cdot\mathbb{P}(Y>d)=e(d)\cdot\mathbb{P}(Y>d).
\end{equation}
If we assume that losses above some threshold $u$ have a $\mathcal{GPD}(u,\sigma,\alpha)$ distribution, for any $d>u$,
\begin{equation}
e_u(d)=\left(\frac{\sigma-u}{\alpha-1} + \frac{\alpha }{\alpha-1} d\right)
\end{equation}
and therefore
\begin{equation}
\pi_u(d)=\frac{n_u}{n}\left[1+ \left(\frac{d-u}{\sigma}\right)\right]^{-\alpha}\cdot \left(\frac{\sigma-u}{\alpha-1} + \frac{\alpha }{\alpha-1} d\right).
\end{equation}
If we plugin estimators of $\sigma$ and $\alpha$, we derive the estimator of the pure premium $\pi_u(d)$. Note that in Section 4.6. in \shortciteN{AlBeTe:17}, approximations are given for Extended Pareto distributions.

As previously, consider a dataset with $n$ observations, i.i.d., $y_1,\cdots,y_n$, assume that above threshold $u$, the (conditional) distribution is $\mathcal{GPD}({u,\sigma,\alpha})$ and let $n_u$ denote the number of observations above $u$. Then if we fit a GPD distribution, the estimated premium would be 
\begin{equation}
\widehat{\pi}_u(d)=\frac{n_u}{n}\left[1+ \left(\frac{d-u}{\widehat{\sigma}}\right)\right]^{-\widehat{\alpha}}\cdot \left(\frac{\widehat{\sigma}-u}{\widehat{\alpha}-1} + \frac{{\alpha} }{\widehat{\alpha}-1} d\right),
\end{equation}
for some deductible $d$, higher than $u$, our predefined threshold.

\subsection{Application on Real Data}
 
\begin{figure}[ht]
\centering\includegraphics[width=.98\textwidth]{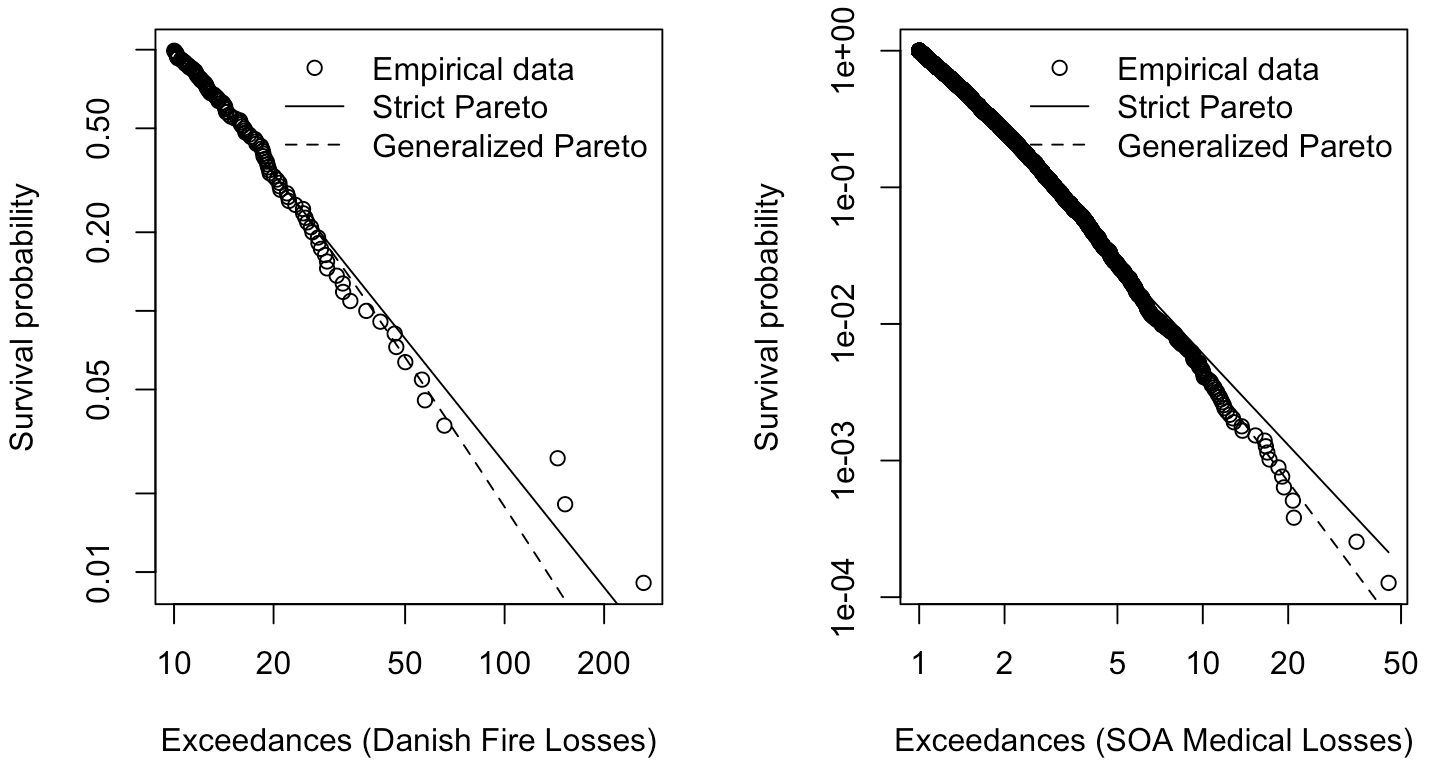}
\caption{Pareto plot, with the Danish Fire data on the left, and the SOA Medical claims data on the right.}\label{fig1a}
\end{figure}

In order to illustrate, two datasets are considered, with a large fire loss dataset (the `Danish dataset' studied in \shortciteNP{BeirlantTeugels1992} and \shortciteNP{McNeil:1997})\footnote{It is the \texttt{danishuni} dataset in the \texttt{CASdatasets} package, available from {\sffamily http://cas.uqam.ca/}} and large medical losses (from the SOA dataset studied in \shortciteNP{cebrian2003}). On Figure \ref{fig1a}, Pareto plots are considered for losses above some threshold $u$, with $u=10$ for fire losses, and $u=1$ for medical losses (here in $\$ '00,000$), i.e.
\begin{equation}
\left(x_{i:n},1-\frac{i-(n-n_u)}{n_u}\right)_{i=n-n_u+1,\cdots,n}
\end{equation}
on a log-log scatterplot. In Pareto models, points should be on a straight line, with slope $-\alpha$. The plain line (\full) is the Pareto distribution (fitted using maximum likelihood techniques) and the dotted line ({\dashed}) corresponds to a Generalized Pareto distribution. 

\begin{figure}[ht]
\centering\includegraphics[width=.48\textwidth]{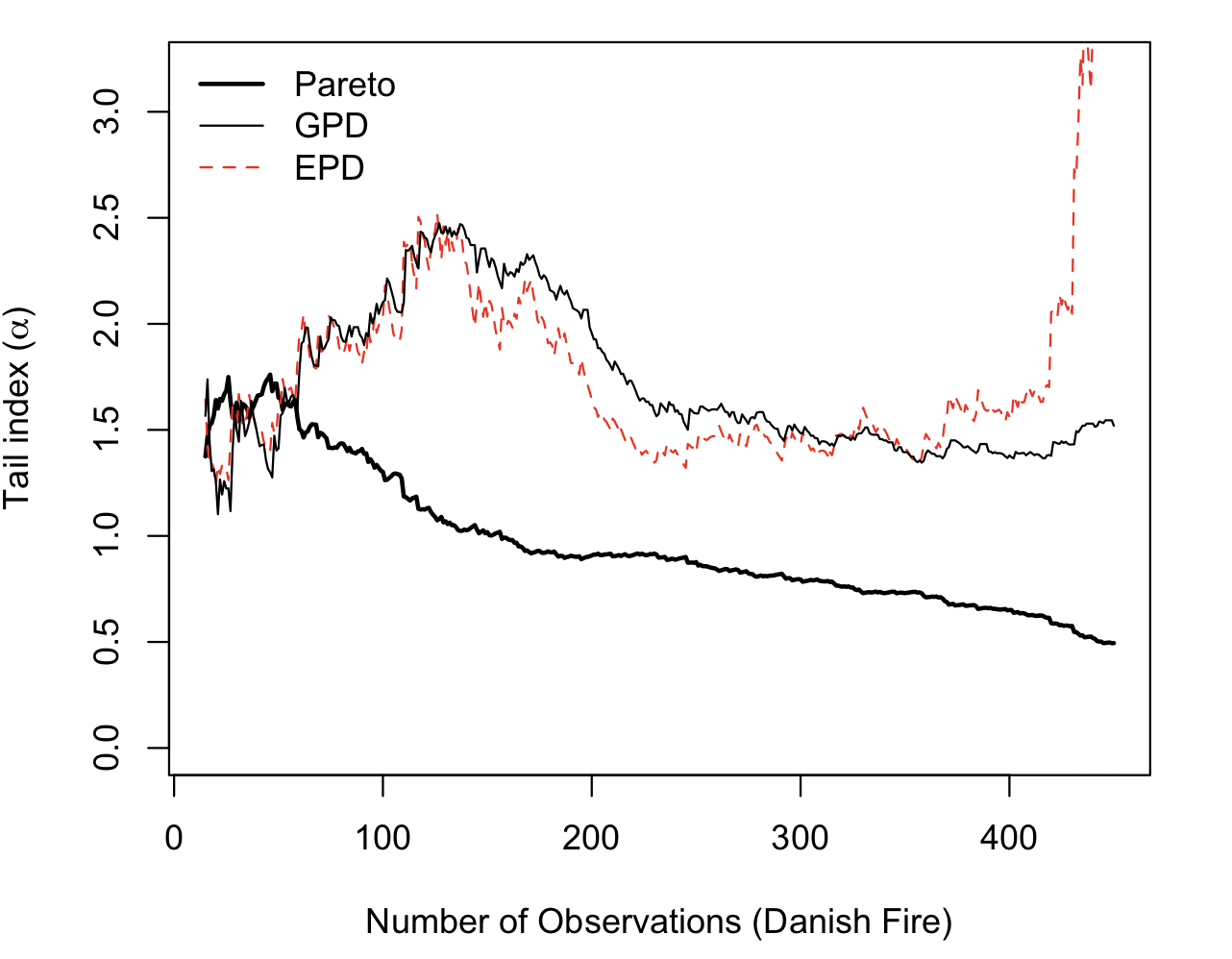}\centering\includegraphics[width=.48\textwidth]{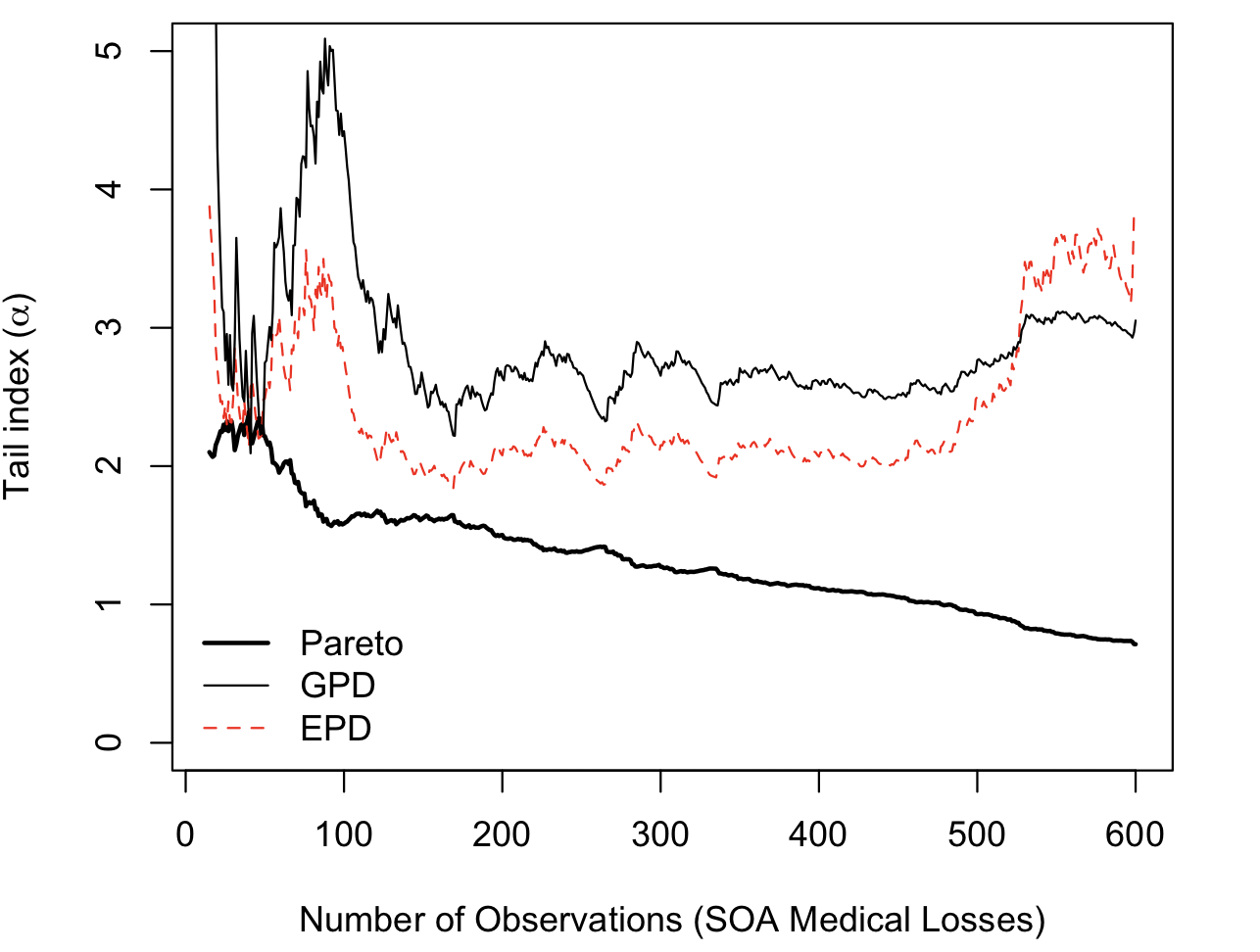}
\caption{Estimation of $\alpha$, with a Pareto model (\fulllwd), a Generalized Pareto model (\full) and an Extended Pareto Model (\textcolor{red}{\dashed}), as a function of $n_u$, the number of tail events considered.}\label{fig1alpha}
\end{figure}

On Figure \ref{fig1alpha} {we can visualize estimates of $\alpha$, as a function of the number of tail events considered}. The strong line (\fulllwd) is the Pareto distribution, the thin line (\full)  is the GPD distribution while the dotted line (\textcolor{red}{\dashed}) corresponds to a Extended Pareto distribution. 
Here the three distributions were fitted using maximum likelihood techniques, and only $\widehat{\alpha}$ is ploted.

\begin{figure}[ht]
\centering\includegraphics[width=.98\textwidth]{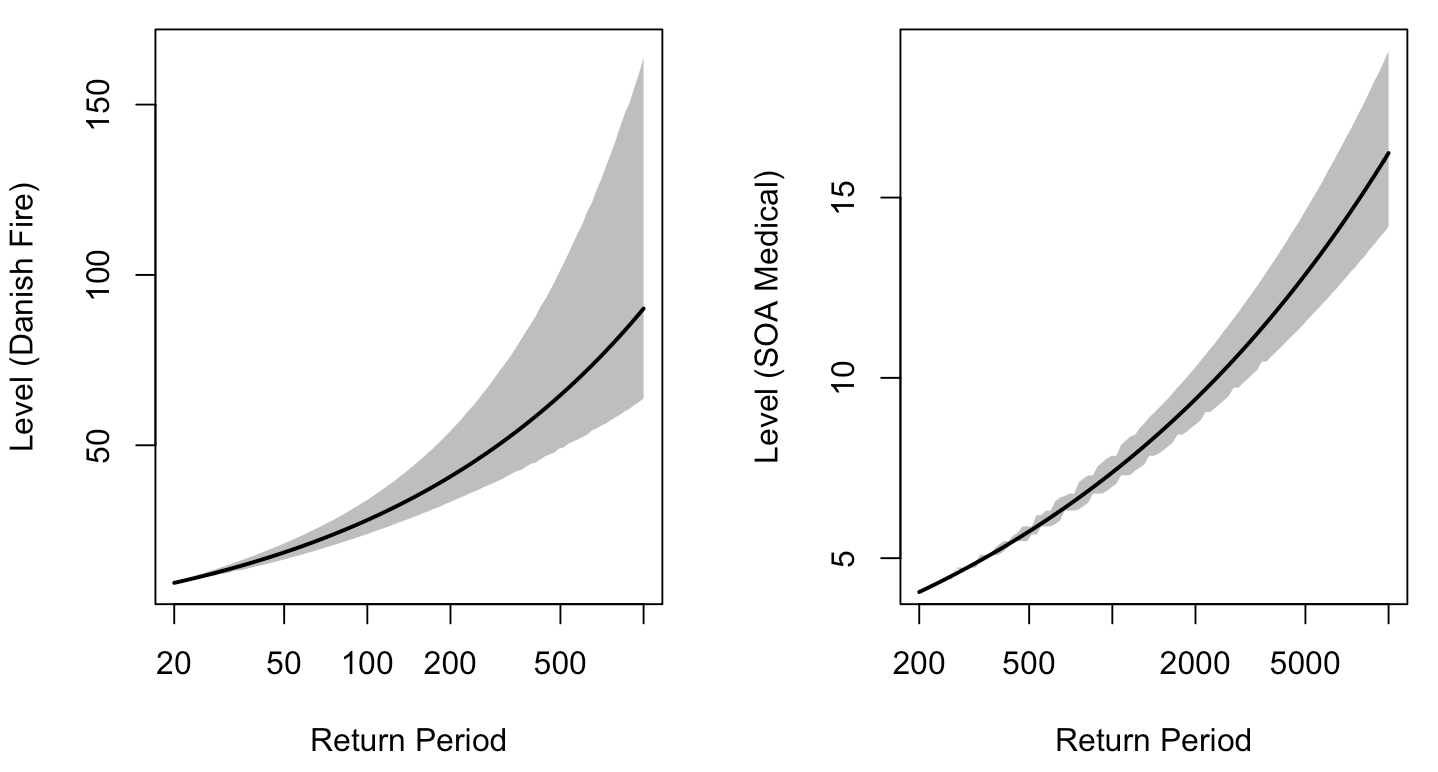}
\caption{Estimation return levels for various return periods, for with the Danish Fire data on the left, and the SOA Medical claims data on the right. The grey area is the confidence region.}\label{fig1b}
\end{figure}

On Figure \ref{fig1b}, we can visualize return levels for various return periods (on a log-scale), when Generalized Pareto distributions are fitted above level $u$ ($u=10$ for fires and $u=2$ for medical claims). Those values are quantiles, then return periods are interpreted as probabilities (e.g. a return period of 200 is a quantile of level $99.5\%$).

\begin{figure}[ht]
\centering\includegraphics[width=.48\textwidth]{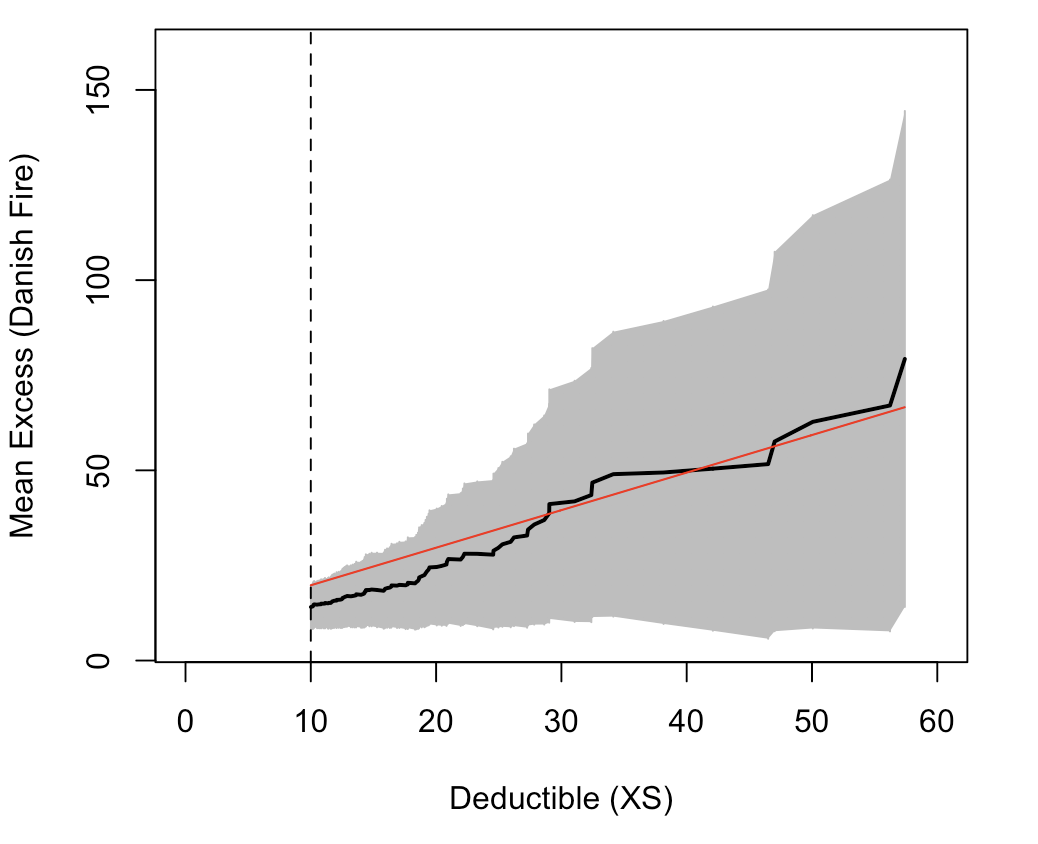}\centering\includegraphics[width=.48\textwidth]{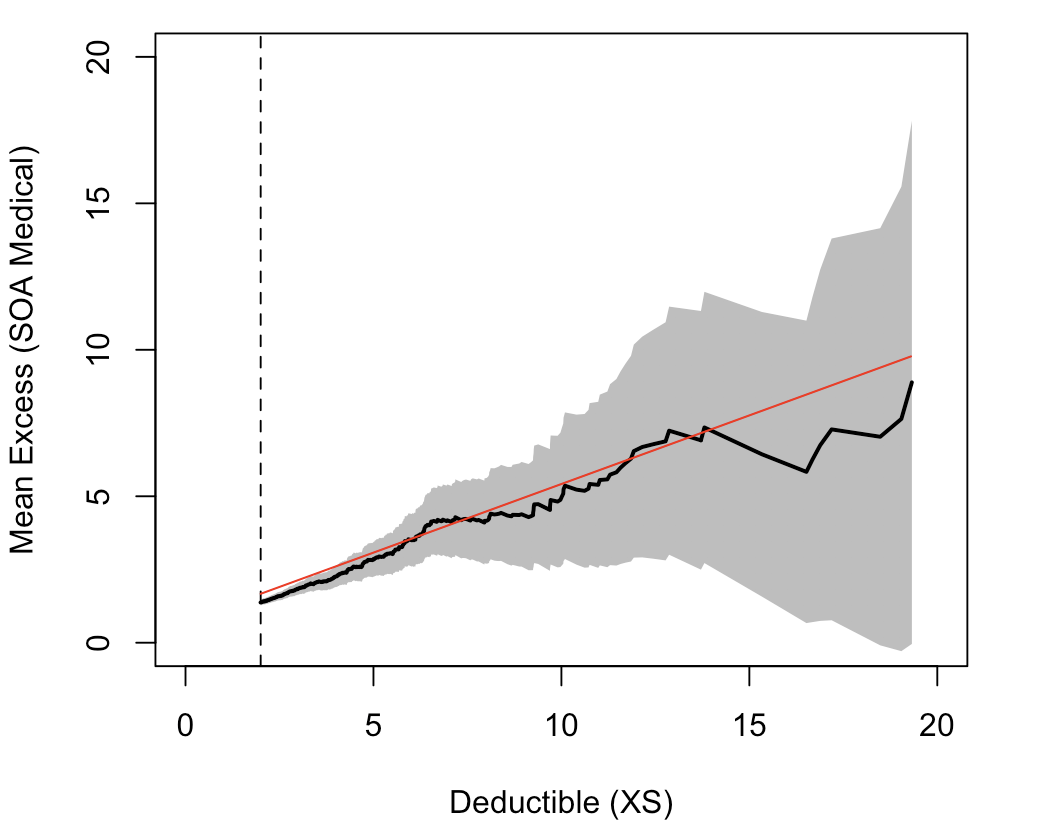}
\caption{Mean excess functions Danish Fire data on the left, and the SOA Medical claims data on the right, with confidence intervals on the empirical version. The line (\textcolor{red}{\full}) is the GPD fit, for a given threshold (respectively 10 and 2).}\label{fig1c}
\end{figure}

On Figure \ref{fig1c}, we can visualize the mean excess function of individual claims, for fire losses on the left, and medical claims on the right (in \$ '00,000). The dark line is the empirical version (average above some threshold) and the line (\textcolor{red}{\full}) is the GPD fit
\begin{equation}
\widehat{e}_u(d)=\left(\frac{\widehat{\sigma}-u}{\widehat{\alpha}-1} + \frac{\widehat{\alpha} }{\widehat{\alpha}-1} d\right)
\end{equation}
respectively with $u=10$ on the left and $u=2$ on the right, for various values of the threshold $d$ ($\geq u$).

\begin{figure}[ht]
\centering\includegraphics[width=.48\textwidth]{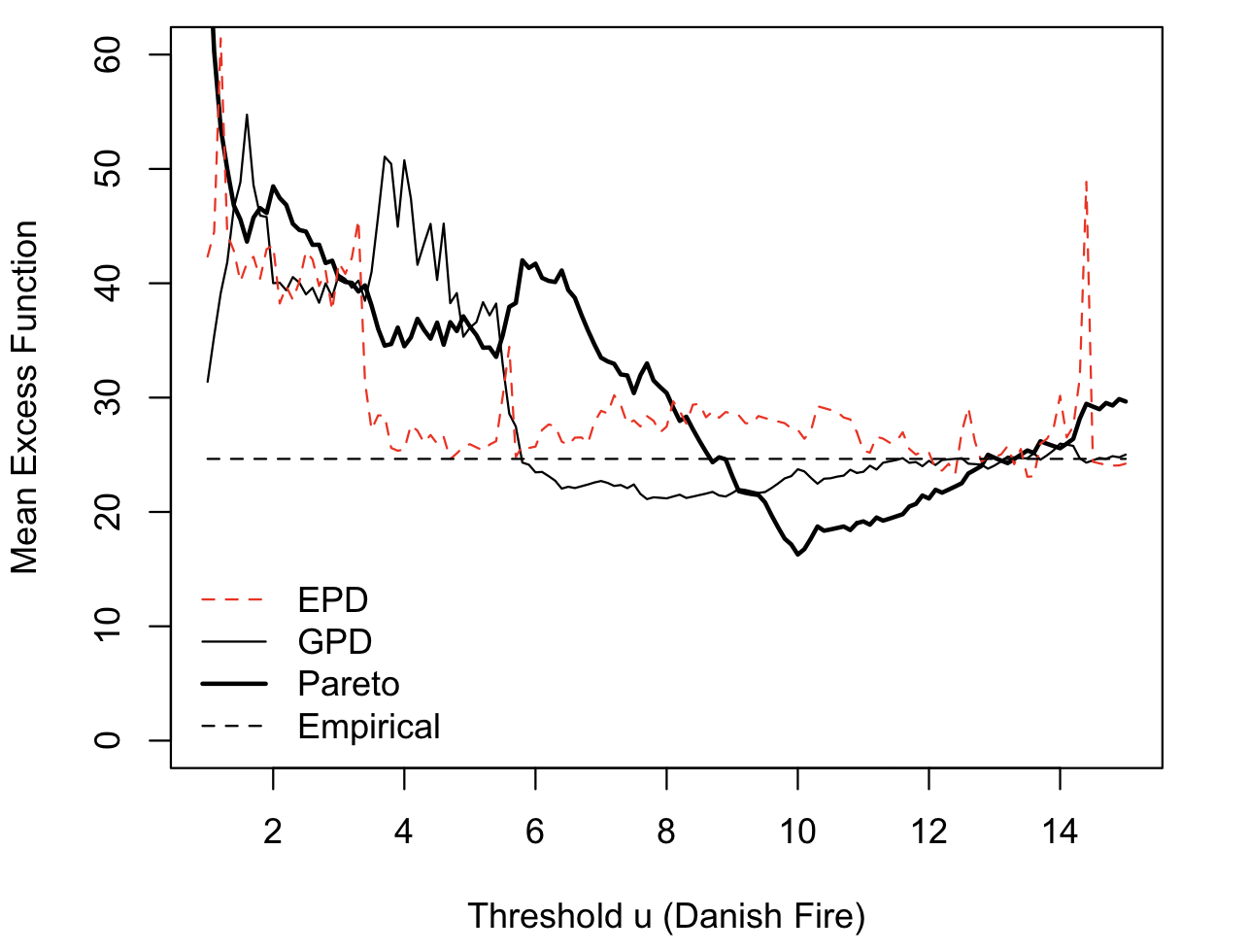}\centering\includegraphics[width=.48\textwidth]{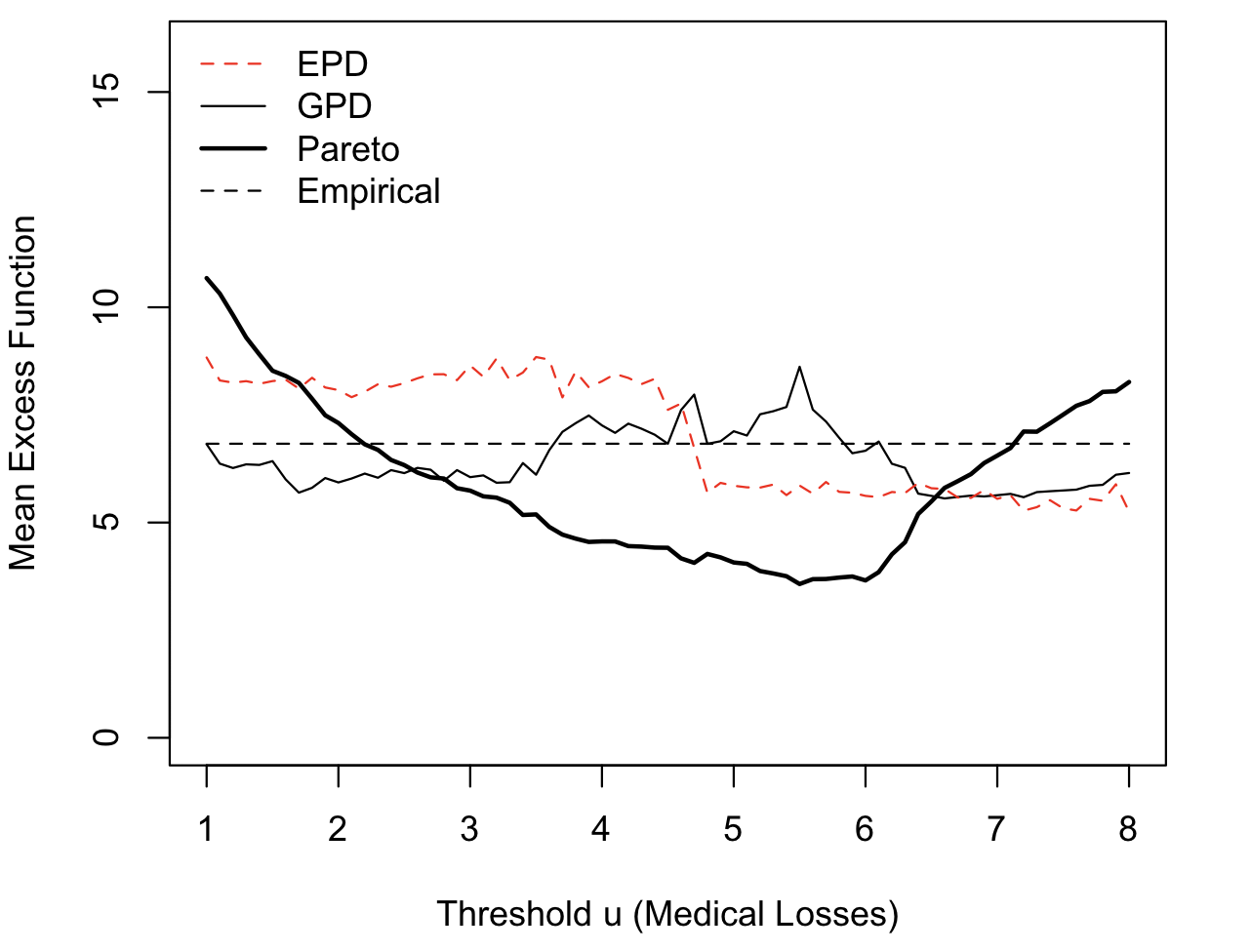}
\caption{Estimation of $e(d)$, with $d=20$ for Danish fires and $d=12$ for Medical losses, with the empirical average ({\dashed}), the estimator derived from a Pareto distribution above threshold $u$ (\fulllwd), from a Generalized Pareto model above threshold $u$ (\full) and from an Extended Pareto Model above threshold $u$ (\textcolor{red}{\dashed}), as a function of $u$.}\label{fig1cevol}
\end{figure}

On Figure \ref{fig1cevol}, we can visualize the stability of estimators with respect to the choice of the threshold $u$, with the estimation of $e(d)$, with $d=20$ for Danish fires and $d=12$ for Medical losses. The empirical average is the horizontal dashed line ({\dashed}), the estimator derived from a Pareto distribution above threshold $u$ is the plain strong line (\fulllwd), the one from a Generalized Pareto model above threshold $u$ is the plain line (\full) and the Extended Pareto model above threshold $u$ is the dashed line (\textcolor{red}{\dashed}), as a function of $u$. The estimator obtained from the Extended Pareto model is quite stable, and close to the empirical estimate.

\begin{figure}[ht]
\centering\includegraphics[width=.48\textwidth]{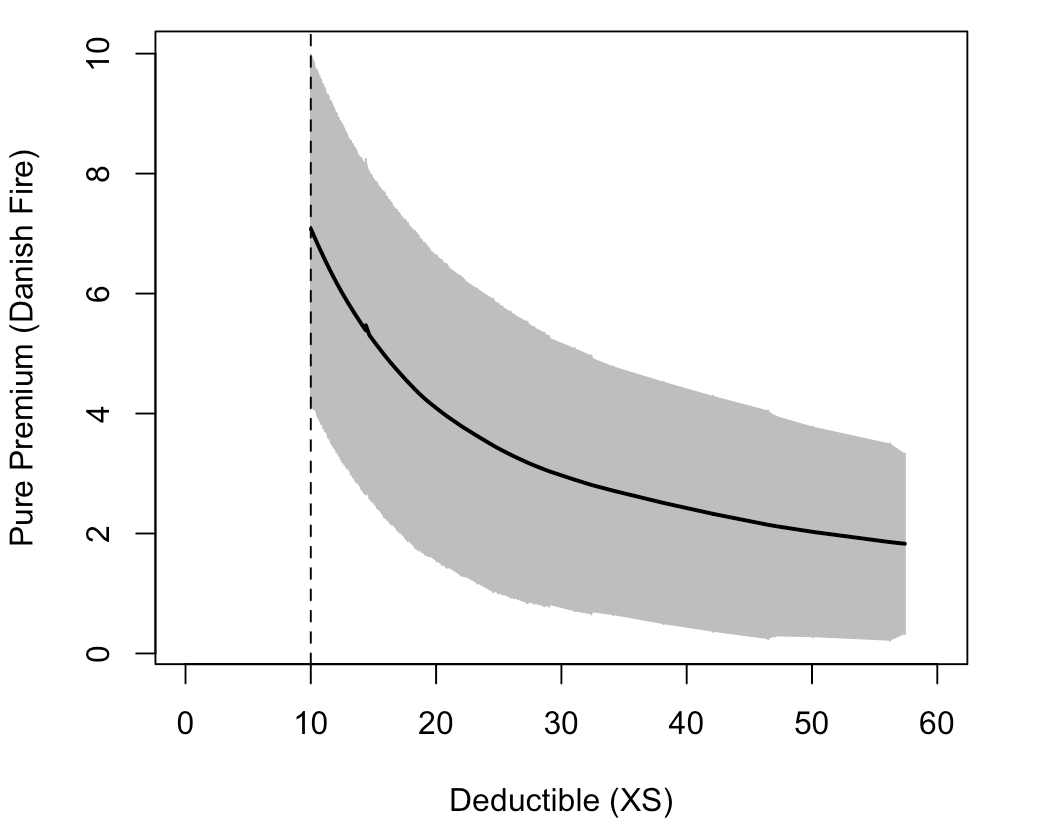}\centering\includegraphics[width=.48\textwidth]{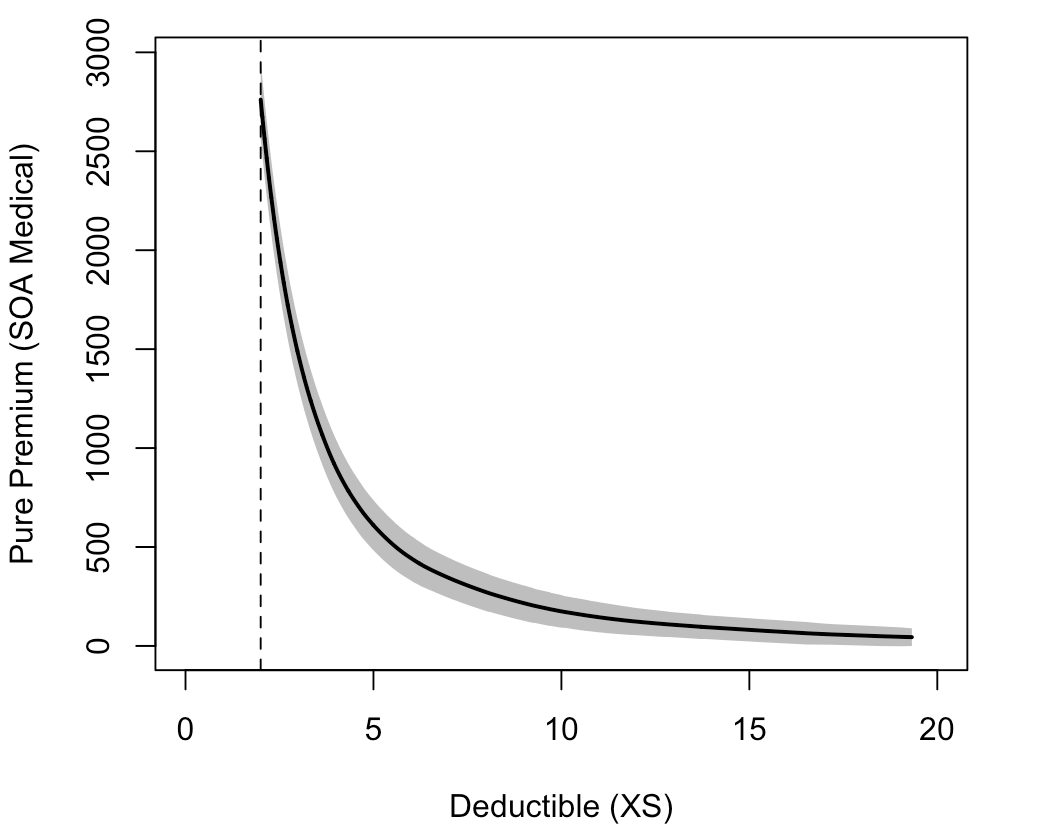}
\caption{Annual Pure Premium for Danish Fire data on the left, and the SOA Medical claims data on the right.}\label{fig1d}
\end{figure}

Finally, on Figure \ref{fig1d}, we have the evolution of the (annual) pure premium, which is simply the mean excess function, multiplied by the (annual) probability to have a loss that exceed the deductible. For the Danish fire losses, since we have 10 years of data, the premium takes into account this 10 factor (which is consistent with a Poisson process assumption for claims occurrence): on average there are 11 claims per year above 10, and 2.5 above 25. For the SOA medical losses, there are 213 claims above \$ 500,000 ($u=5$) for a mean excess function slightly below 3, so the pure premium is close to 600, for a deductible (per claim) of 5.

\section{Finance and risk-measures}\label{sec:7}

\subsection{Downside Risk Measures}

Classical problems in risk management in financial applications involve (extreme) quantile estimation, such as the Value-at-Risk and the Expected Shortfall. Let $Y$ denote the negative returns of some financial instrument, then the Value-at-Risk $\text{VaR}(p)$ is defined as the quantile of level $1-p$ of the distribution,
\begin{equation}
\text{VaR}(p)=Q(1-p)=\inf\big\lbrace y\in\mathbb{R}:F(y)\geq 1-p\big\rbrace,
\end{equation}
while the expected shortfall (or tail conditional expectation) is the potential size of the loss exceeding the Value-at-Risk,
\begin{equation}
\text{ES}(p)=\mathbb{E}\big[X\big\vert X>\text{VaR}(p)\big]
\end{equation}
If $X_t$ is the (random) loss variable of a financial position from time $t$ to time $t+h$ (with a given time horizon $h$), classical distributions have been intensively consider in financial literature, starting from the Gaussian case, $X_t\sim\mathcal{N}(\mu_t,\Sigma_t)$. Then
\begin{equation}
\text{VaR}(p)=\mu_t + \Phi^{-1}(1-p)\Sigma_t \quad\text{and}\quad\text{ES}_p=\mu_t+\frac{\phi(\Phi^{-1}(1-p))}{p}\Sigma_t,
\end{equation}
where $\phi$ and $\Phi$ denote respectively the density and the c.d.f. of the centered standard Gaussian distribution.

Since Pareto models have a simple expression to model losses above a threshold, with a GPD above a threshold $u$, for $y>u$,
\begin{align*}
\widehat{F}_u(y)& = \mathbb{P}[Y\leq y]\\
 &=\mathbb{P}[Y\leq y|Y> u]\cdot \mathbb{P}[Y> u]+\mathbb{P}[Y\leq u]\\
 &=G_{\widehat{\sigma},\widehat{\alpha}}(y-u)\cdot\frac{n_u}{n} + \frac{n-n_u}{n}
\end{align*}
where $n_u$ denotes the number of observations above threshold $u$, and
\begin{equation}
G_{\sigma,\alpha}(x) = 1-\left[1+ \left(\frac{x}{\sigma}\right)\right]^{-\alpha} \qquad\text{for } x\geq u,
\end{equation}
is the $\mathcal{GPD}(0,\sigma,\alpha)$ CDF.
Then for an event rare enough - i.e. a probability small enough ($p<n_u/n$) - the Value-at-Risk $\text{VaR}_p$ can be approximated by
\begin{equation}
\widehat{\text{VaR}}_u(p) = u\cdot \widehat{\sigma}\left[\left(\frac{p}{q_u}\right)^{-1/\widehat{\alpha}}-1\right]^{-1},\text{ where }q_u=\mathbb{P}[Y> u],
\end{equation}
as well as the expected-shortfall,
\begin{equation}
\widehat{\text{ES}}_u(p) =\frac{\widehat{\alpha} \widehat{\text{VaR}}_u(p)+\widehat{\sigma}-u}{\widehat{\alpha}-1}.
\end{equation}

\subsection{Application on Real Data}

In finance, log-returns are rarely independent, and it is necessary to take into account the dynamics of the volatility. Consider some GARCH-type process for log-returns, $y_t =\mu_t+ \Sigma_t x_t$, where $(x_t)$ are i.i.d. variables, and where $(\sigma_t)$ is the stochastic volatility.
Then, as mentioned in \citeN{McNeilFrey2000} 
\begin{eqnarray}
&\widehat{\text{VaR}}_{Y_t|\Sigma_t}(p)=\mu_t+\Sigma_t\cdot \widehat{\text{VaR}}_{X_t}(p) \\ 
&\widehat{\text{ES}}_{Y_t|\Sigma_t}(p)=\mu_t+\Sigma_t\cdot \widehat{\text{ES}}_{X_t}(p)
\end{eqnarray}
To estimate $(\Sigma_t)$, several approaches can be considered, from the exponential weighted moving average (EWMA), with
\begin{equation}
\widehat{\Sigma}_{t+1}^2 = \beta\widehat{\Sigma}_{t}+(1-\beta)y^2_t,\text{ for some }\beta\in(0,1),
\end{equation}
to the GARCH(1,1) process,
\begin{equation}
\widehat{\Sigma}_{t+1}^2 = \alpha_0+\alpha_1 y^2_{t}+ \beta_1\widehat{\Sigma}_{t}\text{ for some }\alpha_1,\beta_1\in(0,1),\text{ with }\alpha_1+\beta_1<1.
\end{equation}

To illustrate, we use returns of the Brent crude price of oil\footnote{Available from {\sffamily https://www.nasdaq.com/market-activity/commodities/bz\%3Anmx}}, extracted from the North Sea prices and comprises Brent Blend, Forties Blend, Oseberg and Ekofisk crudes.

\begin{figure}[ht]
\centering\includegraphics[width=.98\textwidth]{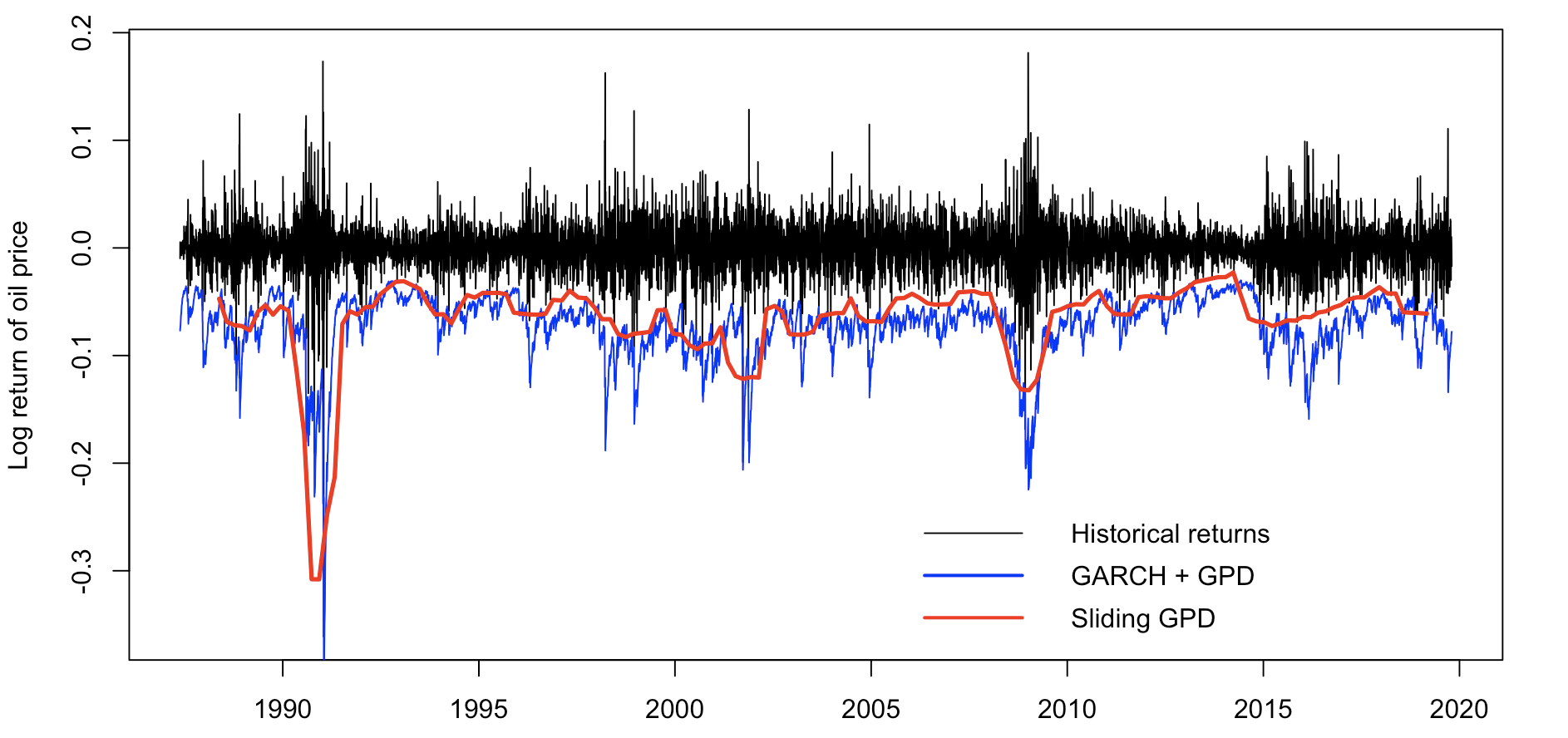}
\caption{Value-at-Risk at level 0.5\% for the daily log-return of oil prices (\full), with a GARCH+GPD model (\textcolor{blue}{\full}) and a sliding window estimator GPD model (\textcolor{red}{\full})).}\label{fig2a}
\end{figure}

On Figure \ref{fig2a}, for the plain line (GARCH+GPD, \textcolor{blue}{\full}) we use 
\begin{equation}
\widehat{\text{VaR}}_{Y_t|\Sigma_t}(p)=\widehat{\Sigma}_t\cdot \widehat{\text{VaR}}_{X_t}(p)
\end{equation}
where 
\begin{equation}
\widehat{\text{VaR}}_{X_t}(p)=u+\frac{\widehat{\sigma}}{\widehat{\alpha}}\left[\left(\frac{n}{n_u}(1-q)\right)^{-\widehat{\alpha}}-1\right]
\end{equation}
with $u=2\%$ and $\widehat{\Sigma}_t$ is the fitted volatility with a GARCH(1,1) process. The red line is obtained using a Generalized Pareto fit on a sliding window $[t\pm 150]$.

\begin{figure}[ht]
\centering\includegraphics[width=.98\textwidth]{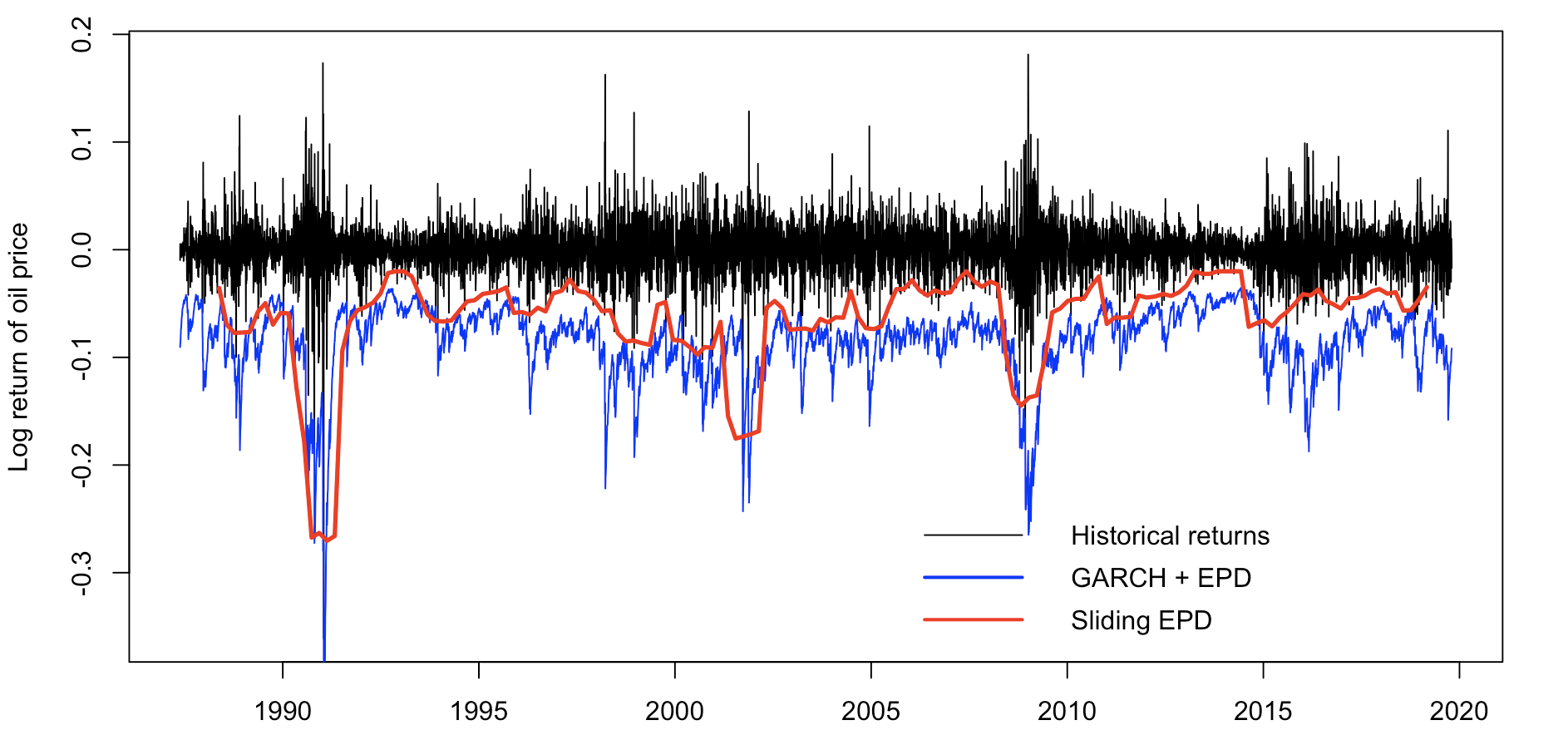}
\caption{Value-at-Risk at level 0.5\% for the daily log-return of oil prices (\full), with a GARCH+EPD model (\textcolor{blue}{\full}) and a sliding window estimator EPD model (\textcolor{red}{\full}).}\label{fig2b}
\end{figure}

On Figure \ref{fig2b}, an extended Pareto model is considered for $\widehat{\text{VaR}}_{X_t}(p)$.

\begin{figure}[ht]
\centering\includegraphics[width=.98\textwidth]{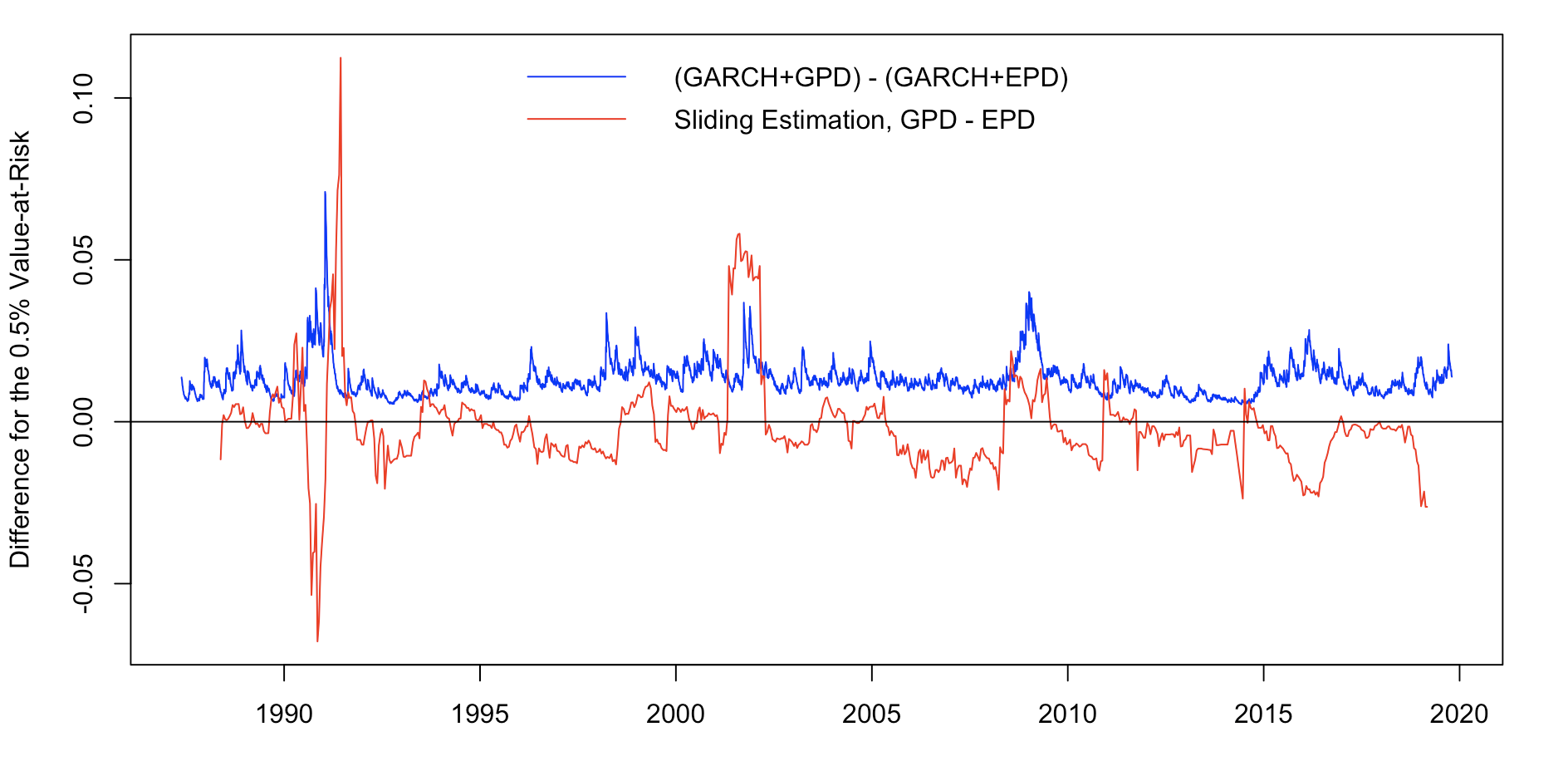}
\caption{Difference of the Value-at-Risk at level 0.5\% for the daily log-return of oil prices between GARCH+Pareto models (\textcolor{blue}{\full}) and a sliding window estimator of Pareto models (\textcolor{red}{\full}).}\label{fig2c}
\end{figure}

Finally, on Figure \ref{fig2c}, we compare the GARCH + Pareto models, with either a Generalized Pareto model (as in \citeN{McNeilFrey2000}) or an Extended Pareto model. Observe that the 0.05\% Value-at-Risk is always over estimated with the Generalized Pareto model. Nevertheless, with Pareto models estimated on sliding windows (on $[t\pm 150]$) the difference can be negative, sometimes. But overall, it is more likely that the Generalized Pareto overestimates the Value-at-Risk.

\bibliographystyle{chicago}
\bibliography{bibliography}

\end{document}